\documentclass[%
 reprint,
amsmath,amssymb,
aps,
prl,
twocolumn,
]{revtex4-2}
\usepackage{braket}
\usepackage{graphicx}
\usepackage{bm}
\usepackage{float}
\usepackage{color}
\usepackage{hyperref}
\usepackage{placeins}

\usepackage[normalem]{ulem}

\usepackage[explicit]{titlesec}
\titleformat{\section}
  {\normalfont}{}{1em}{\noindent \textit{#1} --- }

\begin{document}

\title{Reshaping quantum device noise via repetition code circuits}

\author{Yue Ma$^1$}

\author{Michael Hanks$^1$}

\author{Evdokia Gneusheva$^1$}

\author{M. S. Kim$^1$}

\affiliation{$^1$Blackett Laboratory, Imperial College London, SW7 2AZ, United Kingdom\\
}

\begin{abstract}

Noise of a quantum processor can be an important resource for simulating open quantum dynamics. However, this requires characterizing the device noise and then transforming it into a target structure.
Here we take the first step towards this goal:
We analytically and numerically study reshaping the noise associated with native trapped-ion two-qubit entangling gates via quantum circuits based on repetition codes, and experimentally demonstrate our findings on the IonQ Aria-1 quantum hardware. We investigate all the building blocks, including the quantum channels describing noisy two-qubit entangling gates, the compilation of the encoding circuits into native gates, and the propagation of two-qubit errors across ideal single-qubit gates.

\end{abstract}


\maketitle



All quantum computers~\cite{montanaro2016quantum} suffer from noises -- uncontrollable interactions, some of which lead to undesired evolutions --errors~\cite{preskill2018quantum,fedorov2022quantum} (we use ``noise" and ``error" interchangeably in this manuscript). Two categories of methods have been developed to effectively remove noise: quantum error correction (QEC)~\cite{shor1995scheme,calderbank1996good,steane1996multiple,raussendorf2012key,campbell2017roads,girvin2021introduction,georgescu202025,wallman2016noise,ma2023non,fowler2012surface,shor1996fault,trout2018simulating,debroy2020logical} and quantum error mitigation (QEM)~\cite{cai2023quantum}. However, a question much less explored but more relevant to the current NISQ era is, how the noise can be exploited, one important application being assisting simulations of open quantum dynamics. This application has been demonstrated in the context of QEM~\cite{guimaraes2023noise,ma2024limitations}, and in some theoretically designed algorithms~\cite{leppakangas2023quantum,bertrand2024turning}. But these works mostly assume single-qubit noise models, which do not directly correspond to the noisiest element in realistic quantum devices. The complicated structures of native noise have not been involved in any of these studies.



Here we address this challenge of directly exploiting native device noise. We focus on the trapped-ion quantum devices, where the mediator for qubit-qubit interactions, the common vibrational mode, is also a quantum degree of freedom~\cite{sorensen2000entanglement,molmer1999multiparticle,ma2022unifying}. The two-qubit entangling gates, which are a source of seriously noisy quantum operations, have been experimentally characterized~\cite{wang2020high,fang2022crosstalk,kang2023designing}, but the highlights are not the native noise structure.
In this paper, we first derive this noise structure. Then we demonstrate reshaping it into quantum channels that have different noise structures, analytically, numerically and experimentally, via quantum circuits based on the three-qubit bit-flip 
QEC code. Our choice of QEC circuits does not mean that we aim at correcting errors. Instead, the three-qubit QEC circuit represents one of the smallest meaningful quantum computing units, in contrast to random circuits which are much more widely used but are arguably of limited practical usage~\cite{lubinski2023application}.


\noindent \textit{Noisy M{\o}lmer-S{\o}rensen gates} --- M{\o}lmer-S{\o}rensen (MS) gates~\cite{molmer1999multiparticle,sorensen2000entanglement} implement the two-qubit entangling gates in trapped-ion quantum devices. Qubits (ions' electronic states) effectively interact with each other via their common vibrational mode. The dominant noise for MS gates has been pointed out to be from the motional dephasing~\cite{sun2024quantum}, whose Lindblad form is $C=\sqrt{\Gamma_P}a^{\dagger}a$, with $\Gamma_P$ the dephasing rate, and $\hat{a}$ the annihilation operator for the vibrational mode with frequency $\nu$. The dynamics are analytically solvable under certain \textit{approximations}~\cite{sorensen2000entanglement}: In the interaction picture with respect to all the unitary dynamics, the state of the two qubits is always separable from the state of the vibrational mode, and the latter is kept in a thermal state with mean phonon number $n_{\mathrm{th}}$. For the fully entangling gate which is equivalent to the CNOT gate up to single-qubit gates~\cite{sorensen2000entanglement,maslov2017basic,native_gates}, we find that the canonical Kraus operators of the quantum channel that maps the target state to the state produced by the noisy gate have the following structure,
\begin{align}\label{eq:Kraus_short}
    K_1&=a_1I+a_3J_y^2,\nonumber\\
    K_2&=k_2J_y,\nonumber\\
    K_3&=a_2I+a_4J_y^2.
\end{align}
Here $J_y=(Y_1\otimes I_2+I_1\otimes Y_2)/2$ is the total angular momentum and $Y_i$ are Pauli operators. The coefficients have closed form solutions and in particular, the value of $k_2$ uniquely determines the values of $a_1,a_2,a_3,a_4$ (the derivations and results are in the Supplemental Material~\cite{supp}). Our results of canonical Kraus operators go beyond Ref.~\cite{sorensen2000entanglement} where only the fidelity is derived.

\begin{figure}[t!]
    \centering
    \includegraphics[width=0.48\textwidth]{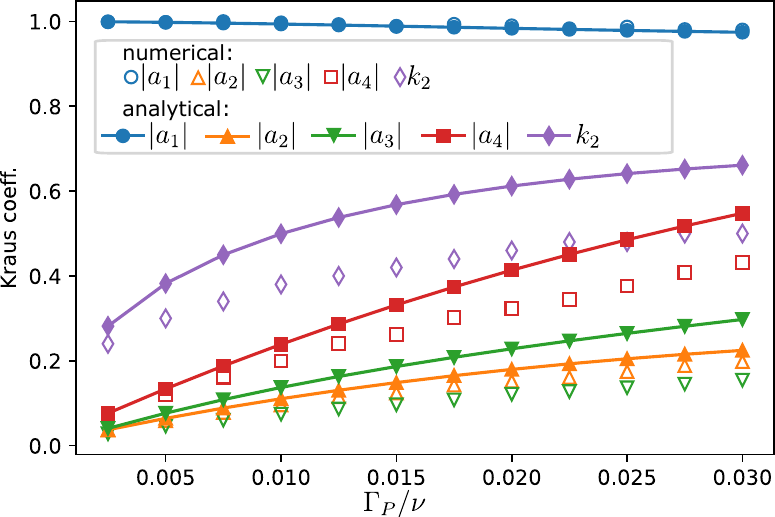}
    \caption{\label{fig:Kraus_coefficients}
    Coefficients of the canonical Kraus operators in Eq.~\eqref{eq:Kraus_short}, as the value of $\Gamma_P$ changes, comparing the analytical approximate solutions (solid line with solid markers) with the numerical exact solutions (open markers).
    }
\end{figure}

An example of how these coefficients change with the dephasing rate is shown in Fig.~\ref{fig:Kraus_coefficients}. We have chosen the parameters~\cite{sorensen2000entanglement} according to the weak coupling regime that most experimental realizations of MS gates follow~\cite{monroe2021programmable,foss2024progress,nam2020ground,wright2019benchmarking}: The Lamb-Dicke parameter is $\eta=0.1$, the Rabi frequency is $\Omega=0.1\nu$, $n_{\mathrm{th}}=0.05$, and $K=25$ is an integer describing the number of loops traversed in the vibrational phase space~\cite{ma2022unifying}. For comparison, in Fig.~\ref{fig:Kraus_coefficients}, we also show numerical simulations obtained without the approximations  above, where we find that the canonical Kraus operators are still in the form of Eq.~\eqref{eq:Kraus_short}, and the coefficients have only reasonably small deviations from the analytical ones. Importantly, $K_2$ is the dominant error term. Here the Pauli Y operator originates from the conventional choice of basis in theoretical derivations~\cite{sorensen2000entanglement}, where the entangling gate is also defined in the $y$-basis. For trapped-ion quantum devices such as IonQ Aria-1, commonly the fully entangling gate is in the $x$-basis~\cite{native_gates}, therefore the dominant error in the noisy MS gate is  $X_1\otimes I_2+I_1\otimes X_2$, i.e, the superposition of the bit-flip error for each qubit. 

Having derived the noise structure of the MS gate, we want to explore how to reshape it.  As the first step, we use one of the smallest meaningful quantum circuit units -- the circuit for the three-qubit bit-flip repetition code. We increase the tunability by adding redundant MS gate pairs and an additional single-qubit gate. One benefit of a QEC circuit is that we map multiple physical qubits into one logical qubit, so that the output has a manageable Hilbert space dimension. We are not actually trying to recover the original state.


\begin{figure}[t!]
    \centering
    \includegraphics[width=0.48\textwidth]{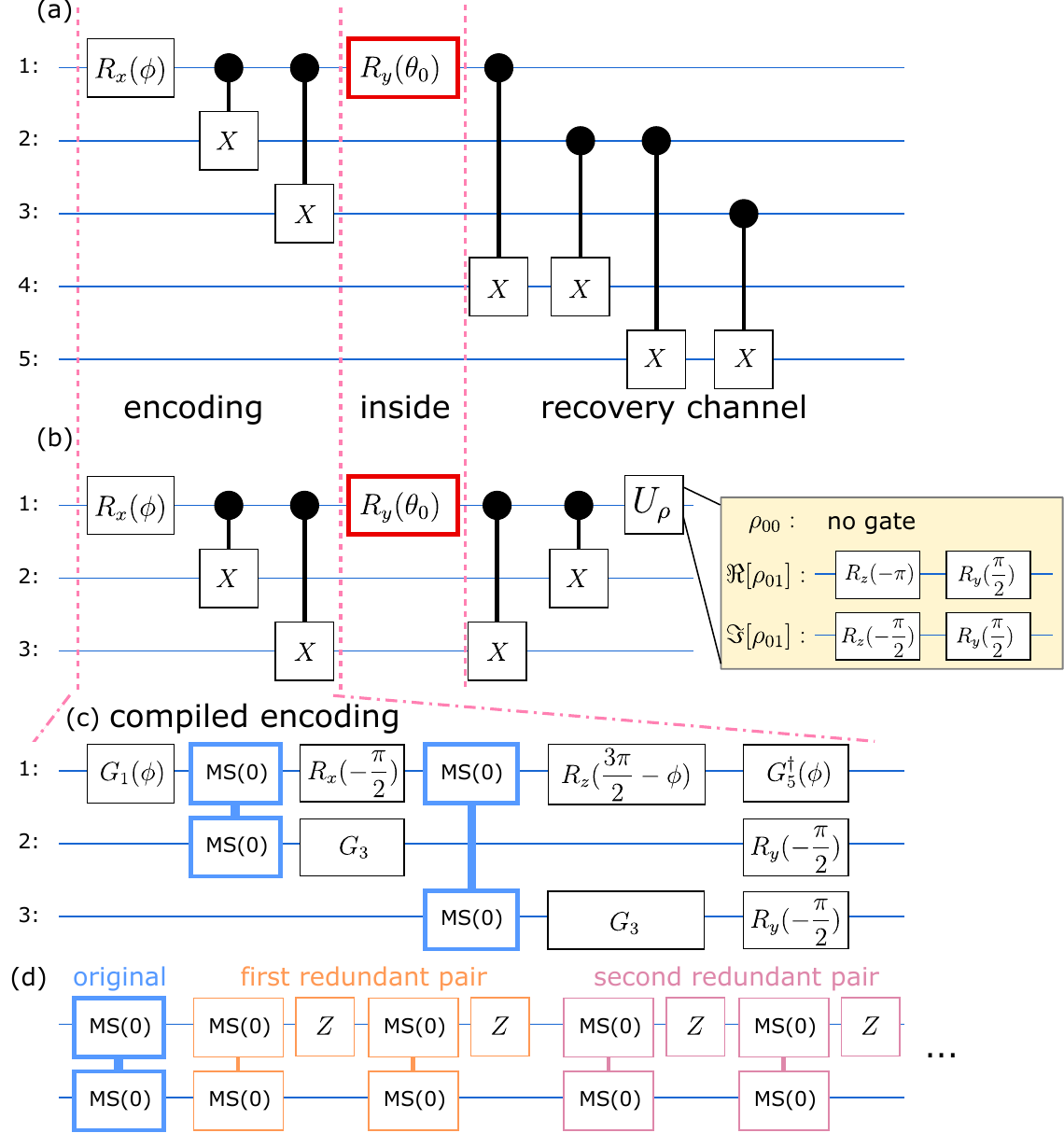}
    \caption{\label{fig:5-qubit-circuit}
    Quantum circuits to implement the modified repetition code for noise reshaping. In (a) and (b), we show the circuits before any compilation into native gates. They are partitioned into three sections as marked. All qubits are measured at the end in the computational basis. The recovery unitaries are implemented via classical post-processing. (a) Two ancilla are included for measuring $\rho_{00}$ with reduced measurement errors. (b) Choosing three different $U_{\rho}$ corresponds to measuring $\rho_{00}$, $\Re[\rho_{01}]$ or $\Im[\rho_{01}]$. (c) The encoding section is first compiled by converting CNOT gates to the MS gates. (d) Redundant pairs of MS gates can be inserted after each encoding MS gate to effectively tune the noise level.
    }
\end{figure}

\noindent \textit{Noisy encoding through bit-flip repetition code} --- QEC codes work by encoding quantum information into a larger Hilbert space such that detection of errors will not completely collapse the state, thus allowing recovery operations after syndrome measurements. For the three-qubit bit-flip repetition code, a state $|\psi\rangle=\alpha|0\rangle+\beta|1\rangle$ is mapped into a three-qubit state, $|\bar{\psi}\rangle=\alpha|000\rangle+\beta|111\rangle$. Each of the three qubits is called a physical qubit, while $|\bar{\psi}\rangle$ defines a logical qubit: The logical space is spanned by $|\bar{0}\rangle=|000\rangle$ and $|\bar{1}\rangle=|111\rangle$. The most straightforward way of implementing this encoding process is via the first section of the quantum circuits in Fig.~\ref{fig:5-qubit-circuit}(a,b). We use  $R_x(\phi)$ to initialize $|\psi\rangle$ with $\alpha=\cos(\phi/2)$ and $\beta=-i\sin(\phi/2)$. The subsequent two CNOT gates encode this into the logical state $|\bar{\psi}\rangle$. For standard QEC theoretical study, this encoding process is considered as noiseless while single-qubit error happens after encoding. However, it is not what we consider.

In order to be implemented on a real trapped-ion quantum device, this encoding circuit section needs to be compiled into an equivalent version consisted of native gates. Of particular interest are the two-qubit entangling gates, as they are noisier than single-qubit gates. Moreover, we are interested in reshaping the native noise of two-qubit gates into quantum channels that have different types of noise. The native MS gate is related to the CNOT gate by applying additional single qubit gates (see the Supplemental Material~\cite{supp}, where all the conventions are also clarified). Compiling the two encoding CNOT gates into two MS gates results in the compiled encoding circuit section as shown in Fig.~\ref{fig:5-qubit-circuit}(c). There the single-qubit gates are defined as $G_1(\phi)=-\frac{1}{\sqrt{2}}\cos\frac{\phi}{2}I+\frac{i}{\sqrt{2}}\sin\frac{\phi}{2}X+\frac{i}{\sqrt{2}}\cos\frac{\phi}{2}Y-\frac{i}{\sqrt{2}}\sin\frac{\phi}{2}Z$, $G_3=\frac{1}{2}I+\frac{i}{2}X-\frac{i}{2}Y-\frac{i}{2}Z$ and $G_5^{\dagger}(\phi)=\frac{1}{\sqrt{2}}\cos\frac{\phi}{2}I+\frac{i}{\sqrt{2}}\cos\frac{\phi}{2}X-\frac{i}{\sqrt{2}}\sin\frac{\phi}{2}Y-\frac{i}{\sqrt{2}}\sin\frac{\phi}{2}Z$. Although this is not the full compilation which should include converting all the single-qubit gates into trapped-ion native gates and is shown in the Supplemental Material~\cite{supp} for the case of IonQ Aria-1 quantum device, this intermediate circuit is illustrative for keeping track of the propagation of two-qubit noise while assuming ideal single-qubit gates, which we will focus on later.

We have derived that the noisy implementation of each MS gate is represented by applying the quantum channel with Kraus operators Eq.~\eqref{eq:Kraus_short} (replacing the $y$-axis with $x$) after the noiseless MS gate. For a given quantum device, the amount of error is fixed and roughly corresponds to the left side of Fig.~\ref{fig:Kraus_coefficients}. As our main interest is about how to reshape this native noise, we would like to have tunability over the strength of MS gate noise. We achieve this by adding redundant pairs of MS gates~\cite{wang2020high,fang2022crosstalk} after each of the two encoding MS gates, as shown in Fig.~\ref{fig:5-qubit-circuit}(d). Each pair consists of the following sequence repeated two times: one MS gate followed by one $Z$ gate on the first qubit. In the noiseless case each redundant pair equals to identity. In practice, adding redundant pairs changes the noise strength and structure associated with the circuit, which will be discussed later. Note that, the noise in implementing the MS gates already brings the state of the three physical qubits out of the logical space.

\noindent \textit{Additional tunability inside the code} --- We define ``inside the code" as operations after the encoding circuit section and before the recovery channel (defined later). Traditional QEC studies assume that errors happen here, i.e., bit-flip repetition code is to correct single-qubit stochastic bit-flip error $X$. However, as we do not want to inject noise that is not native, we instead add tunability by inserting a parameterized single-qubit gate. One of the simplest options is to add a unitary gate associated with rotation around the $y$-axis~\cite{braketConvention},
\begin{equation}\label{eq:Ry}
R_y(\theta_0)=
    \begin{pmatrix}
        \cos(\theta_0/2) & -\sin(\theta_0/2)\\
        \sin(\theta_0/2) & \cos(\theta_0/2)
    \end{pmatrix},\ \ 0\leq \theta_0\leq\pi,
\end{equation}
on physical qubit 1 (see the Supplemental Material~\cite{supp} for more generalizations). This is illustrated as the red box in Fig.~\ref{fig:5-qubit-circuit}(a,b). Adding this $R_y$ gate may bring the three-qubit state further away from the logical space. This may thus be viewed as a coherent error from the standard QEC perspective. It is not how we interpret it here. We are interested in how the code reshapes the noise profile of quantum computers, rather than the classical number of the QEC success rate: The output logical state from the repetition code can be viewed as the initial logical state going through a quantum channel that is different from the one on the physical qubit level. We will next elaborate it by explicitly describing how the recovery channel bring the state back to the logical space.

\noindent \textit{Recovery channel to the logical space} --- As per the standard definition of the three-qubit bit-flip repetition code, we perform syndrome measurements on the stabilizers $(Z_1 Z_2, Z_2 Z_3)$, which are expressed as projectors, and apply unitary operators conditioning on the measurement results. This process is formally defined as the recovery channel $\mathcal{R}(\tilde{\rho})=\sum_{i,j=0}^1R_{1+2i+j}P^{(ij)}\tilde{\rho} P^{(ij)}R_{1+2i+j}$, where the projectors are $P^{(ij)}=(I+(-1)^iZ_1Z_2+(-1)^jZ_2Z_3+(-1)^{i+j}Z_1Z_3)/4$, with index $0(1)$ meaning the syndrome measurement result is $+1(-1)$, and the applied unitaries are $(R_1,R_2,R_3,R_4)=(I,X_3,X_1,X_2)$. The recovery channel maps any three-qubit state $\tilde{\rho}$ into the output logical state $\bar{\rho}_f$ in the logical code space. Here $\tilde{\rho}$ is the outcome of applying the qubit 1 $R_y(\theta_0)$ gate onto the state resulting from the noisy encoding circuit. The latter regarding error propagation will be explained later.

In Fig.~\ref{fig:5-qubit-circuit}(a,b), we show the practical implementations of the recovery channel. Note that we assume perfect single-qubit gates and two-qubit gates in this section of the circuits as we do not introduce any redundant gates in contrast to what we do in the encoding section. Considering the limitations of quantum devices, we do not explicitly apply mid-circuit measurements or multi-qubit operators. Instead we measure all qubits in the computational basis at the end of the circuits and perform classical post-processing. The circuit in Fig.~\ref{fig:5-qubit-circuit}(a) explicitly uses two ancilla qubits (labelled 4 and 5) to compensate for measurement errors: Only results that are compatible with the code space basis $\{|000\rangle_{123},|111\rangle_{123}\}$ are counted in the statistics. The circuit in Fig.~\ref{fig:5-qubit-circuit}(b) does not include ancilla, but is able to estimate both $\langle000|\bar{\rho}_f|000\rangle\equiv\rho_{00}$ and the complex-valued $\langle000|\bar{\rho}_f|111\rangle\equiv\rho_{01}$ once the corresponding single-qubit gate $U_{\rho}$ is selected at the end. More details are in the Supplemental Material~\cite{supp}.

\noindent \textit{Errors associated with noisy native gates} --- The noisy implementation of each MS gate can be approximated keeping the first order as mapping any state $\rho$ into
\begin{equation}
    \mathrm{MS}_{ij}\rho\mathrm{MS}_{ij}^{\dagger}+\sqrt{\frac{r_1}{2}}(X_i+X_j)\mathrm{MS}_{ij}\rho\mathrm{MS}_{ij}^{\dagger}(X_i+X_j)\sqrt{\frac{r_1}{2}},\label{eq:MSnoise}
\end{equation}
where we approximate $a_1\approx1$, $k_2\approx\sqrt{2r_1}$, and $a_2\approx a_3\approx a_4\approx0$, and have replaced $y$ with $x$ based on Eq.~\eqref{eq:Kraus_short}, and $i,j$ are labels indicating which two qubits the MS gate acts on. Note the invariance of exchanging $i$ and $j$.

Appending redundant MS gate pairs to one original MS gate as illustrated in Fig.~\ref{fig:5-qubit-circuit}(d) changes the structure of errors. By keeping track of how $X_{i,j}$ terms propagate across $Z$ gates, we find that the unit in Fig.~\ref{fig:5-qubit-circuit}(d) with $n$ pairs of redundant MS gates corresponds to mapping any state $\rho$ into
\begin{align}
    &\mathrm{MS}_{ij}\rho\mathrm{MS}_{ij}^{\dagger}\nonumber\\
    +&\sqrt{\frac{r_1n}{2}}(-X_i+X_j)\mathrm{MS}_{ij}\rho\mathrm{MS}_{ij}^{\dagger}(-X_i+X_j)\sqrt{\frac{r_1n}{2}}\nonumber\\
    +&\sqrt{\frac{r_1(n+1)}{2}}(X_i+X_j)\mathrm{MS}_{ij}\rho\mathrm{MS}_{ij}^{\dagger}(X_i+X_j)\sqrt{\frac{r_1(n+1)}{2}}.\label{eq:MSnoiseRepeat}
\end{align}
The strength of noise is indeed enhanced by a factor of $n$. However, the difference of bit-flip error now contributes together with the existing sum of bit-flip error: Both indicate coherent superposition of single-qubit errors. 

Further error propagation towards the end of the encoding circuit section will further transform the error structures. We find that, if we insert $n$ redundant pairs of MS gates after each of the two encoding MS gates, the physical state of the three qubits at the end of the encoding circuit section is approximated by
\begin{align}
    |\bar{\psi}\rangle\langle\bar{\psi}|&+\frac{r_1(n+1)}{2}(Z_1+X_2)|\bar{\psi}\rangle\langle\bar{\psi}|(Z_1+X_2)\nonumber\\
    &+\frac{r_1(n+1)}{2}(Z_1+X_3)|\bar{\psi}\rangle\langle\bar{\psi}|(Z_1+X_3)\nonumber\\
    &+\frac{r_1n}{2}(-Z_1+X_2)|\bar{\psi}\rangle\langle\bar{\psi}|(-Z_1+X_2)\nonumber\\
    &+\frac{r_1n}{2}(-Z_1+X_3)|\bar{\psi}\rangle\langle\bar{\psi}|(-Z_1+X_3),\label{eq:state_approx}
\end{align}
where $|\bar{\psi}\rangle$ is the ideal encoded initial state.The existence of single-qubit gates has converted some bit-flip errors into phase-flip errors. 

\noindent \textit{Noise reshaping into the logical channel} --- The state as approximately expressed in Eq.~\eqref{eq:state_approx} is then acted upon by the tunable $R_y$ gate inside the code as marked in Fig.~\ref{fig:5-qubit-circuit}(a,b). Then the recovery channel is implemented. This channel maps any three-qubit physical state into the logical space (also defined as the code space). The error structure contained in this output logical state is the result of noise reshaping that we are interested in. To be specific, we have shown that $\bar{\rho}_f=\rho_{00}|\bar{0}\rangle\langle\bar{0}|+\rho_{01}|\bar{0}\rangle\langle\bar{1}|+\rho_{01}^*|\bar{1}\rangle\langle\bar{0}|+(1-\rho_{00})|\bar{1}\rangle\langle\bar{1}|$. This state contains the same quantum information as the single-qubit state $\rho_f=\rho_{00}|0\rangle\langle0|+\rho_{01}|0\rangle\langle1|+\rho_{01}^*|1\rangle\langle0|+(1-\rho_{00})|1\rangle\langle1|$. The quantum channel that maps the initial state $|\psi\rangle$ into $\rho_f$ is the result of reshaping the native device noise: the noise associated with MS gates, as shown in Eq.~\eqref{eq:Kraus_short}, has a fundamentally different structure.

\begin{figure}[t!]
    \centering
    \includegraphics[width=0.48\textwidth]{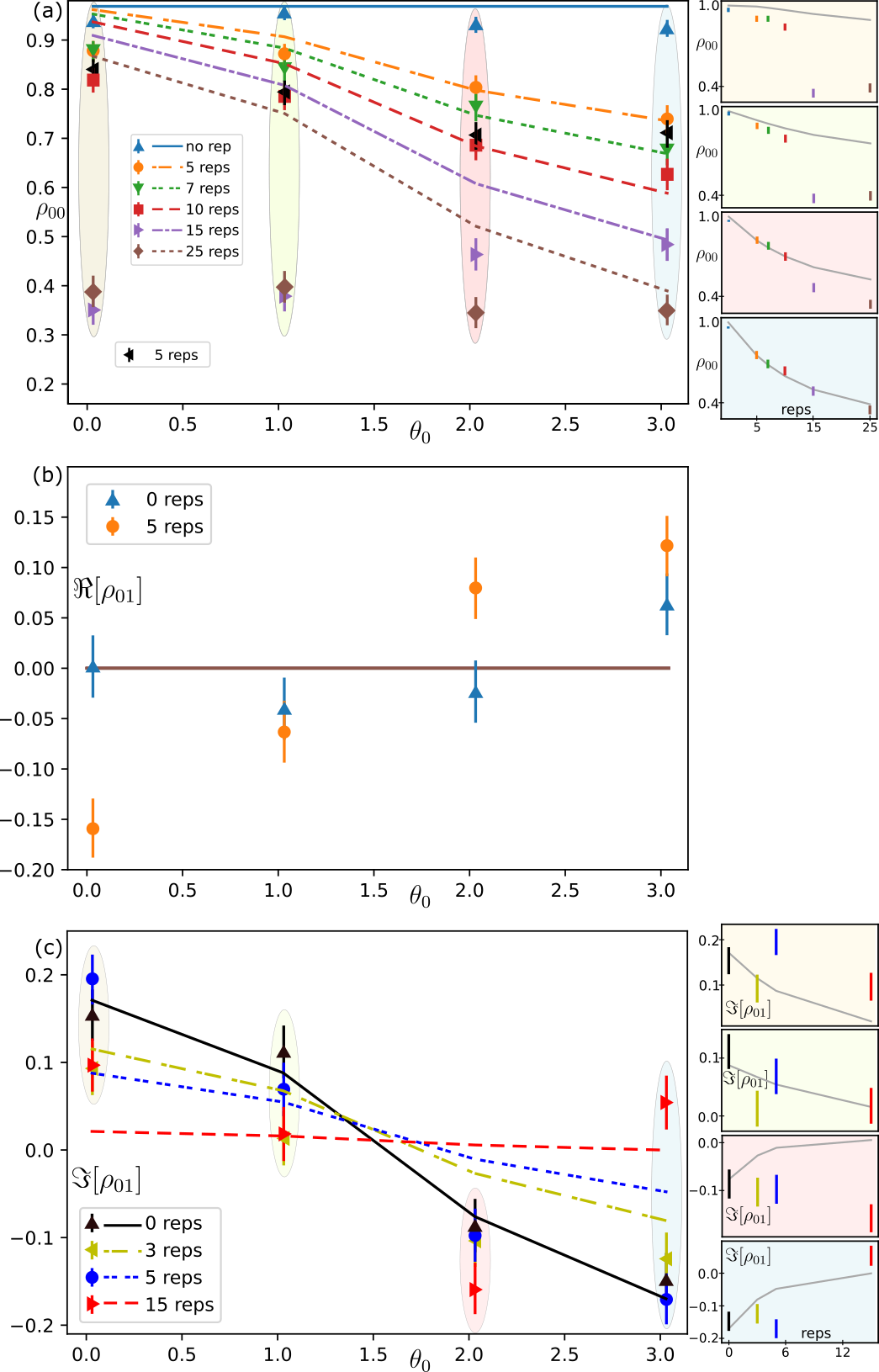}
    \caption{\label{fig:data_rho00}
    Data obtained from the IonQ Aria-1 quantum device (points), in comparison with our model (lines). Each data point is averaged over ${1000}$ shots. Error bars are ${95\%}$ confidence intervals computed from the Wilson score \cite{wilson_probable_1927}. The number of repetitions refers to the number of the inserted pairs of redundant MS gates after each encoding gate. (a) ${\rho_{00}}$. The data come from the compiled version of Fig.~\ref{fig:5-qubit-circuit}(a), except the black left-pointing triangles which are from the compiled circuit of Fig.~\ref{fig:5-qubit-circuit}(b) for reference (see the main text for details). (b\{c\}) The real \{imaginary\} part of $\rho_{01}$. The data come from the compiled circuit of Fig.~\ref{fig:5-qubit-circuit}(b).}
    
\end{figure}

We can derive the analytical form of $\bar{\rho}_f$ following the linear approximation of two-qubit entangling gate errors as shown in Eq.~\eqref{eq:MSnoise} and Eq.~\eqref{eq:MSnoiseRepeat}. However, for the effective noise to be large enough to be resolvable numerically and experimentally, the number of pairs of redundant gates needs to be large enough, i.e., $n\gg1$. This in fact breaks down the linear approximation. Attempting to generalize Eq.~\eqref{eq:MSnoise} and Eq.~\eqref{eq:MSnoiseRepeat} beyond the linear approximation will require keeping track of terms polynomial or even exponential in $n$ -- either way intractable for $n\gg5$. We did not choose to proceed in this way because this exponential dependence is an artifact resulting from our naive way of tuning the error rate by introducing redundancies. It is an open question as to how to efficiently modify the effective error rate of the native gates. We thus numerically calculate $\bar{\rho}_f$ based on the full Kraus operator description of noisy MS gates. Specially, we use the numerically simulated result of the coefficients in Eq.~\eqref{eq:Kraus_short} that corresponds to the left-most values in Fig.~\ref{fig:Kraus_coefficients} (exact numbers see the Supplemental Material~\cite{supp}), which is compatible with the current IonQ Aria-1 fidelity of two-qubit entangling gates. The results are shown as lines in Fig.~\ref{fig:data_rho00}, which will be compared with the data we get from the IonQ Aria-1 quantum device. 


\noindent \textit{IonQ Aria-1 results} --- It is possible to further compile the circuit Fig.~\ref{fig:5-qubit-circuit}(c) into circuits entirely made of IonQ Aria-1 native gates without swapping the positions of the encoding MS gates with any other gate (see the Supplemental Material~\cite{supp}). The results obtained from the quantum device, for using the compiled version of circuit Fig.~\ref{fig:5-qubit-circuit}(a) to estimate $\rho_{00}$, are shown as the data points in Fig.~\ref{fig:data_rho00}(a). Without any redundant pair of MS gates, the dependence of $\rho_{00}$ on $\theta_0$ is weak. Once we add 5, 7 or 10 redundant pairs, it becomes clear that $\rho_{00}$ decreases as $\theta_0$ increases from $0$ to $\pi$. We can also see the decrease of $\rho_{00}$ as the number of redundant pairs increase (see the insets). If we add 15 or 25 redundant pairs, the dependence of $\rho_{00}$ on $\theta_0$ becomes less clear. Altogether we can see that our model (plotted as lines) matches reasonably well with the data, given that we do not run any direct device characterization.

The results from the compiled version of the circuits in Fig.~\ref{fig:5-qubit-circuit}(b), which reconstruct $\rho_{00}$, $\Re[\rho_{01}]$ or $\Im[\rho_{01}]$ without using ancilla qubits, are included in Fig.~\ref{fig:data_rho00}(a-c). In Fig.~\ref{fig:data_rho00}(a), the black data points (left-pointing triangles) are based on the compiled version of Fig.~\ref{fig:5-qubit-circuit}(b) for 5 repetitions. The smaller $\rho_{00}$ compared with the orange circles for each $\theta_0$ results from the lack of post-selection for compensating measurement errors as there is no ancilla qubit. In Fig.~\ref{fig:data_rho00}(b), we compare the data for $\Re[\rho_{01}]$ with our model which predicts $\Re[\rho_{01}]=0$ for all $\theta_0$ and $p$. The data indeed show small values of $\Re[\rho_{01}]$ around 0. The deviations from $0$ can be associated with noise sources other than the MS gate noise. Reading out $\Re[\rho_{01}]$ does not reveal any features from the latter, thus various effects from the former are unveiled. In Fig.~\ref{fig:data_rho00}(c), we compare the data for $\Im[\rho_{01}]$ with our model. For 0, 3 or 5 repetitive MS gate pairs, increasing $\theta_0$ reduces $\Im[\rho_{01}]$, as predicted by our model (lines). However, for each fixed $\theta_0$, how $\Im[\rho_{01}]$ changes with the number of repetitions cannot be fully resolved. This is because $\Im[\rho_{01}]-\theta_0$ curves for different $p$ are close to each other and cross each other, as demonstrated in the lines in Fig.~\ref{fig:data_rho00}(c). In fact, this crossing is visible in the hardware data. For the third value of $\theta_0$, $\Im[\rho_{01}]$ for 0, 3 and 5 repetitions are the closest to each other compared with other $\theta_0$. For 15 repetitive MS gate pairs, the deviation from our model is more evident. We have also included the insets to illustrate the dependence of $\Im[\rho_{01}]$ on the number of redundant pairs $n$ while fixing $\theta_0$.

\noindent \textit{Conclusion and discussion} --- We demonstrate exploiting the native structure of quantum device noise and reshaping it. Focusing on trapped-ion setups, we analytically characterize noisy entangling gates, and investigate how the native noises propagate across and get transformed by modified repetition code circuits. And we verify our results experimentally via the IonQ Aria-1 quantum device. Our work opens up a new research direction fundamentally different from the current mainstream effort on making qubits and gates with higher fidelity. We suggest to focus on exploiting the benefits brought forward by the physics behind the noise, a promising application of which is simulating open quantum dynamics on quantum devices which is very difficult under existing approaches.



\noindent \textit{Acknowledgement} --- We thank Hyukjoon Kwon for very helpful discussions. We acknowledge financial support from EPSRC EP/Y004752/1, EP/W032643/1, EP/Z53318X/1. This work was also supported by the National
Research Foundation of Korea (NRF) grant funded by
the Korea government (MSIT) (No. RS-2024-00413957).
%


\begin{thebibliography}{38}%
\makeatletter
\providecommand \@ifxundefined [1]{%
 \@ifx{#1\undefined}
}%
\providecommand \@ifnum [1]{%
 \ifnum #1\expandafter \@firstoftwo
 \else \expandafter \@secondoftwo
 \fi
}%
\providecommand \@ifx [1]{%
 \ifx #1\expandafter \@firstoftwo
 \else \expandafter \@secondoftwo
 \fi
}%
\providecommand \natexlab [1]{#1}%
\providecommand \enquote  [1]{``#1''}%
\providecommand \bibnamefont  [1]{#1}%
\providecommand \bibfnamefont [1]{#1}%
\providecommand \citenamefont [1]{#1}%
\providecommand \href@noop [0]{\@secondoftwo}%
\providecommand \href [0]{\begingroup \@sanitize@url \@href}%
\providecommand \@href[1]{\@@startlink{#1}\@@href}%
\providecommand \@@href[1]{\endgroup#1\@@endlink}%
\providecommand \@sanitize@url [0]{\catcode `\\12\catcode `\$12\catcode `\&12\catcode `\#12\catcode `\^12\catcode `\_12\catcode `\%12\relax}%
\providecommand \@@startlink[1]{}%
\providecommand \@@endlink[0]{}%
\providecommand \url  [0]{\begingroup\@sanitize@url \@url }%
\providecommand \@url [1]{\endgroup\@href {#1}{\urlprefix }}%
\providecommand \urlprefix  [0]{URL }%
\providecommand \Eprint [0]{\href }%
\providecommand \doibase [0]{https://doi.org/}%
\providecommand \selectlanguage [0]{\@gobble}%
\providecommand \bibinfo  [0]{\@secondoftwo}%
\providecommand \bibfield  [0]{\@secondoftwo}%
\providecommand \translation [1]{[#1]}%
\providecommand \BibitemOpen [0]{}%
\providecommand \bibitemStop [0]{}%
\providecommand \bibitemNoStop [0]{.\EOS\space}%
\providecommand \EOS [0]{\spacefactor3000\relax}%
\providecommand \BibitemShut  [1]{\csname bibitem#1\endcsname}%
\let\auto@bib@innerbib\@empty
\bibitem [{\citenamefont {Montanaro}(2016)}]{montanaro2016quantum}%
  \BibitemOpen
  \bibfield  {author} {\bibinfo {author} {\bibfnamefont {A.}~\bibnamefont {Montanaro}},\ }\href@noop {} {\bibfield  {journal} {\bibinfo  {journal} {npj Quantum Inf.}\ }\textbf {\bibinfo {volume} {2}},\ \bibinfo {pages} {1} (\bibinfo {year} {2016})}\BibitemShut {NoStop}%
\bibitem [{\citenamefont {Preskill}(2018)}]{preskill2018quantum}%
  \BibitemOpen
  \bibfield  {author} {\bibinfo {author} {\bibfnamefont {J.}~\bibnamefont {Preskill}},\ }\href@noop {} {\bibfield  {journal} {\bibinfo  {journal} {Quantum}\ }\textbf {\bibinfo {volume} {2}},\ \bibinfo {pages} {79} (\bibinfo {year} {2018})}\BibitemShut {NoStop}%
\bibitem [{\citenamefont {Fedorov}\ \emph {et~al.}(2022)\citenamefont {Fedorov}, \citenamefont {Gisin}, \citenamefont {Beloussov},\ and\ \citenamefont {Lvovsky}}]{fedorov2022quantum}%
  \BibitemOpen
  \bibfield  {author} {\bibinfo {author} {\bibfnamefont {A.~K.}\ \bibnamefont {Fedorov}}, \bibinfo {author} {\bibfnamefont {N.}~\bibnamefont {Gisin}}, \bibinfo {author} {\bibfnamefont {S.~M.}\ \bibnamefont {Beloussov}},\ and\ \bibinfo {author} {\bibfnamefont {A.~I.}\ \bibnamefont {Lvovsky}},\ }\href@noop {} {\bibfield  {journal} {\bibinfo  {journal} {arXiv:2203.17181}\ } (\bibinfo {year} {2022})}\BibitemShut {NoStop}%
\bibitem [{\citenamefont {Shor}(1995)}]{shor1995scheme}%
  \BibitemOpen
  \bibfield  {author} {\bibinfo {author} {\bibfnamefont {P.~W.}\ \bibnamefont {Shor}},\ }\href@noop {} {\bibfield  {journal} {\bibinfo  {journal} {Phys. Rev. A}\ }\textbf {\bibinfo {volume} {52}},\ \bibinfo {pages} {R2493} (\bibinfo {year} {1995})}\BibitemShut {NoStop}%
\bibitem [{\citenamefont {Calderbank}\ and\ \citenamefont {Shor}(1996)}]{calderbank1996good}%
  \BibitemOpen
  \bibfield  {author} {\bibinfo {author} {\bibfnamefont {A.~R.}\ \bibnamefont {Calderbank}}\ and\ \bibinfo {author} {\bibfnamefont {P.~W.}\ \bibnamefont {Shor}},\ }\href@noop {} {\bibfield  {journal} {\bibinfo  {journal} {Phys. Rev. A}\ }\textbf {\bibinfo {volume} {54}},\ \bibinfo {pages} {1098} (\bibinfo {year} {1996})}\BibitemShut {NoStop}%
\bibitem [{\citenamefont {Steane}(1996)}]{steane1996multiple}%
  \BibitemOpen
  \bibfield  {author} {\bibinfo {author} {\bibfnamefont {A.}~\bibnamefont {Steane}},\ }\href@noop {} {\bibfield  {journal} {\bibinfo  {journal} {Proc. R. Soc. Lond. A: Math. Phys. Eng. Sci. }\ }\textbf {\bibinfo {volume} {452}},\ \bibinfo {pages} {2551} (\bibinfo {year} {1996})}\BibitemShut {NoStop}%
\bibitem [{\citenamefont {Raussendorf}(2012)}]{raussendorf2012key}%
  \BibitemOpen
  \bibfield  {author} {\bibinfo {author} {\bibfnamefont {R.}~\bibnamefont {Raussendorf}},\ }\href@noop {} {\bibfield  {journal} {\bibinfo  {journal} {Philos. Trans. R. Soc. A}\ }\textbf {\bibinfo {volume} {370}},\ \bibinfo {pages} {4541} (\bibinfo {year} {2012})}\BibitemShut {NoStop}%
\bibitem [{\citenamefont {Campbell}\ \emph {et~al.}(2017)\citenamefont {Campbell}, \citenamefont {Terhal},\ and\ \citenamefont {Vuillot}}]{campbell2017roads}%
  \BibitemOpen
  \bibfield  {author} {\bibinfo {author} {\bibfnamefont {E.~T.}\ \bibnamefont {Campbell}}, \bibinfo {author} {\bibfnamefont {B.~M.}\ \bibnamefont {Terhal}},\ and\ \bibinfo {author} {\bibfnamefont {C.}~\bibnamefont {Vuillot}},\ }\href@noop {} {\bibfield  {journal} {\bibinfo  {journal} {Nature}\ }\textbf {\bibinfo {volume} {549}},\ \bibinfo {pages} {172} (\bibinfo {year} {2017})}\BibitemShut {NoStop}%
\bibitem [{\citenamefont {Girvin}(2021)}]{girvin2021introduction}%
  \BibitemOpen
  \bibfield  {author} {\bibinfo {author} {\bibfnamefont {S.~M.}\ \bibnamefont {Girvin}},\ }\href@noop {} {\bibfield  {journal} {\bibinfo  {journal} {arXiv:2111.08894}\ } (\bibinfo {year} {2021})}\BibitemShut {NoStop}%
\bibitem [{\citenamefont {Georgescu}(2020)}]{georgescu202025}%
  \BibitemOpen
  \bibfield  {author} {\bibinfo {author} {\bibfnamefont {I.}~\bibnamefont {Georgescu}},\ }\href@noop {} {\bibfield  {journal} {\bibinfo  {journal} {Nat. Rev. Phys.}\ }\textbf {\bibinfo {volume} {2}},\ \bibinfo {pages} {519} (\bibinfo {year} {2020})}\BibitemShut {NoStop}%
\bibitem [{\citenamefont {Wallman}\ and\ \citenamefont {Emerson}(2016)}]{wallman2016noise}%
  \BibitemOpen
  \bibfield  {author} {\bibinfo {author} {\bibfnamefont {J.~J.}\ \bibnamefont {Wallman}}\ and\ \bibinfo {author} {\bibfnamefont {J.}~\bibnamefont {Emerson}},\ }\href@noop {} {\bibfield  {journal} {\bibinfo  {journal} {Phys. Rev. A}\ }\textbf {\bibinfo {volume} {94}},\ \bibinfo {pages} {052325} (\bibinfo {year} {2016})}\BibitemShut {NoStop}%
\bibitem [{\citenamefont {Ma}\ \emph {et~al.}(2023)\citenamefont {Ma}, \citenamefont {Hanks},\ and\ \citenamefont {Kim}}]{ma2023non}%
  \BibitemOpen
  \bibfield  {author} {\bibinfo {author} {\bibfnamefont {Y.}~\bibnamefont {Ma}}, \bibinfo {author} {\bibfnamefont {M.}~\bibnamefont {Hanks}},\ and\ \bibinfo {author} {\bibfnamefont {M. S.}~\bibnamefont {Kim}},\ }\href@noop {} {\bibfield  {journal} {\bibinfo  {journal} {Phys. Rev. Lett.}\ }\textbf {\bibinfo {volume} {131}},\ \bibinfo {pages} {200602} (\bibinfo {year} {2023})}\BibitemShut {NoStop}%
\bibitem [{\citenamefont {Fowler}\ \emph {et~al.}(2012)\citenamefont {Fowler}, \citenamefont {Mariantoni}, \citenamefont {Martinis},\ and\ \citenamefont {Cleland}}]{fowler2012surface}%
  \BibitemOpen
  \bibfield  {author} {\bibinfo {author} {\bibfnamefont {A.~G.}\ \bibnamefont {Fowler}}, \bibinfo {author} {\bibfnamefont {M.}~\bibnamefont {Mariantoni}}, \bibinfo {author} {\bibfnamefont {J.~M.}\ \bibnamefont {Martinis}},\ and\ \bibinfo {author} {\bibfnamefont {A.~N.}\ \bibnamefont {Cleland}},\ }\href@noop {} {\bibfield  {journal} {\bibinfo  {journal} {Phys. Rev. A}\ }\textbf {\bibinfo {volume} {86}},\ \bibinfo {pages} {032324} (\bibinfo {year} {2012})}\BibitemShut {NoStop}%
\bibitem [{\citenamefont {Shor}(1996)}]{shor1996fault}%
  \BibitemOpen
  \bibfield  {author} {\bibinfo {author} {\bibfnamefont {P.~W.}\ \bibnamefont {Shor}},\ }in\ \href@noop {} {\emph {\bibinfo {booktitle} {Proceedings of 37th conference on foundations of computer science}}}\ (\bibinfo {organization} {IEEE},\ \bibinfo {year} {1996})\ pp.\ \bibinfo {pages} {56--65}\BibitemShut {NoStop}%
\bibitem [{\citenamefont {Trout}\ \emph {et~al.}(2018)\citenamefont {Trout}, \citenamefont {Li}, \citenamefont {Guti{\'e}rrez}, \citenamefont {Wu}, \citenamefont {Wang}, \citenamefont {Duan},\ and\ \citenamefont {Brown}}]{trout2018simulating}%
  \BibitemOpen
  \bibfield  {author} {\bibinfo {author} {\bibfnamefont {C.~J.}\ \bibnamefont {Trout}}, \bibinfo {author} {\bibfnamefont {M.}~\bibnamefont {Li}}, \bibinfo {author} {\bibfnamefont {M.}~\bibnamefont {Guti{\'e}rrez}}, \bibinfo {author} {\bibfnamefont {Y.}~\bibnamefont {Wu}}, \bibinfo {author} {\bibfnamefont {S.-T.}\ \bibnamefont {Wang}}, \bibinfo {author} {\bibfnamefont {L.}~\bibnamefont {Duan}},\ and\ \bibinfo {author} {\bibfnamefont {K.~R.}\ \bibnamefont {Brown}},\ }\href@noop {} {\bibfield  {journal} {\bibinfo  {journal} {New J. Phys.}\ }\textbf {\bibinfo {volume} {20}},\ \bibinfo {pages} {043038} (\bibinfo {year} {2018})}\BibitemShut {NoStop}%
\bibitem [{\citenamefont {Debroy}\ \emph {et~al.}(2020)\citenamefont {Debroy}, \citenamefont {Li}, \citenamefont {Huang},\ and\ \citenamefont {Brown}}]{debroy2020logical}%
  \BibitemOpen
  \bibfield  {author} {\bibinfo {author} {\bibfnamefont {D.~M.}\ \bibnamefont {Debroy}}, \bibinfo {author} {\bibfnamefont {M.}~\bibnamefont {Li}}, \bibinfo {author} {\bibfnamefont {S.}~\bibnamefont {Huang}},\ and\ \bibinfo {author} {\bibfnamefont {K.~R.}\ \bibnamefont {Brown}},\ }\href@noop {} {\bibfield  {journal} {\bibinfo  {journal} {Quantum Sci. Technol.}\ }\textbf {\bibinfo {volume} {5}},\ \bibinfo {pages} {034002} (\bibinfo {year} {2020})}\BibitemShut {NoStop}%
\bibitem [{\citenamefont {Cai}\ \emph {et~al.}(2023)\citenamefont {Cai}, \citenamefont {Babbush}, \citenamefont {Benjamin}, \citenamefont {Endo}, \citenamefont {Huggins}, \citenamefont {Li}, \citenamefont {McClean},\ and\ \citenamefont {O’Brien}}]{cai2023quantum}%
  \BibitemOpen
  \bibfield  {author} {\bibinfo {author} {\bibfnamefont {Z.}~\bibnamefont {Cai}}, \bibinfo {author} {\bibfnamefont {R.}~\bibnamefont {Babbush}}, \bibinfo {author} {\bibfnamefont {S.~C.}\ \bibnamefont {Benjamin}}, \bibinfo {author} {\bibfnamefont {S.}~\bibnamefont {Endo}}, \bibinfo {author} {\bibfnamefont {W.~J.}\ \bibnamefont {Huggins}}, \bibinfo {author} {\bibfnamefont {Y.}~\bibnamefont {Li}}, \bibinfo {author} {\bibfnamefont {J.~R.}\ \bibnamefont {McClean}},\ and\ \bibinfo {author} {\bibfnamefont {T.~E.}\ \bibnamefont {O’Brien}},\ }\href@noop {} {\bibfield  {journal} {\bibinfo  {journal} {Rev. Mod. Phys.}\ }\textbf {\bibinfo {volume} {95}},\ \bibinfo {pages} {045005} (\bibinfo {year} {2023})}\BibitemShut {NoStop}%
\bibitem [{\citenamefont {Guimar{\~a}es}\ \emph {et~al.}(2023)\citenamefont {Guimar{\~a}es}, \citenamefont {Lim}, \citenamefont {Vasilevskiy}, \citenamefont {Huelga},\ and\ \citenamefont {Plenio}}]{guimaraes2023noise}%
  \BibitemOpen
  \bibfield  {author} {\bibinfo {author} {\bibfnamefont {J.~D.}\ \bibnamefont {Guimar{\~a}es}}, \bibinfo {author} {\bibfnamefont {J.}~\bibnamefont {Lim}}, \bibinfo {author} {\bibfnamefont {M.~I.}\ \bibnamefont {Vasilevskiy}}, \bibinfo {author} {\bibfnamefont {S.~F.}\ \bibnamefont {Huelga}},\ and\ \bibinfo {author} {\bibfnamefont {M.~B.}\ \bibnamefont {Plenio}},\ }\href@noop {} {\bibfield  {journal} {\bibinfo  {journal} {PRX Quantum}\ }\textbf {\bibinfo {volume} {4}},\ \bibinfo {pages} {040329} (\bibinfo {year} {2023})}\BibitemShut {NoStop}%
\bibitem [{\citenamefont {Ma}\ and\ \citenamefont {Kim}(2024)}]{ma2024limitations}%
  \BibitemOpen
  \bibfield  {author} {\bibinfo {author} {\bibfnamefont {Y.}~\bibnamefont {Ma}}\ and\ \bibinfo {author} {\bibfnamefont {M. S.}~\bibnamefont {Kim}},\ }\href@noop {} {\bibfield  {journal} {\bibinfo  {journal} {Phys. Rev. A}\ }\textbf {\bibinfo {volume} {109}},\ \bibinfo {pages} {012431} (\bibinfo {year} {2024})}\BibitemShut {NoStop}%
\bibitem [{\citenamefont {Lepp{\"a}kangas}\ \emph {et~al.}(2023)\citenamefont {Lepp{\"a}kangas}, \citenamefont {Vogt}, \citenamefont {Fratus}, \citenamefont {Bark}, \citenamefont {Vaitkus}, \citenamefont {Stadler}, \citenamefont {Reiner}, \citenamefont {Zanker},\ and\ \citenamefont {Marthaler}}]{leppakangas2023quantum}%
  \BibitemOpen
  \bibfield  {author} {\bibinfo {author} {\bibfnamefont {J.}~\bibnamefont {Lepp{\"a}kangas}}, \bibinfo {author} {\bibfnamefont {N.}~\bibnamefont {Vogt}}, \bibinfo {author} {\bibfnamefont {K.~R.}\ \bibnamefont {Fratus}}, \bibinfo {author} {\bibfnamefont {K.}~\bibnamefont {Bark}}, \bibinfo {author} {\bibfnamefont {J.~A.}\ \bibnamefont {Vaitkus}}, \bibinfo {author} {\bibfnamefont {P.}~\bibnamefont {Stadler}}, \bibinfo {author} {\bibfnamefont {J.-M.}\ \bibnamefont {Reiner}}, \bibinfo {author} {\bibfnamefont {S.}~\bibnamefont {Zanker}},\ and\ \bibinfo {author} {\bibfnamefont {M.}~\bibnamefont {Marthaler}},\ }\href@noop {} {\bibfield  {journal} {\bibinfo  {journal} {Phys. Rev. A}\ }\textbf {\bibinfo {volume} {108}},\ \bibinfo {pages} {062424} (\bibinfo {year} {2023})}\BibitemShut {NoStop}%
\bibitem [{\citenamefont {Bertrand}\ \emph {et~al.}(2024)\citenamefont {Bertrand}, \citenamefont {Besserve}, \citenamefont {Ferrero},\ and\ \citenamefont {Ayral}}]{bertrand2024turning}%
  \BibitemOpen
  \bibfield  {author} {\bibinfo {author} {\bibfnamefont {C.}~\bibnamefont {Bertrand}}, \bibinfo {author} {\bibfnamefont {P.}~\bibnamefont {Besserve}}, \bibinfo {author} {\bibfnamefont {M.}~\bibnamefont {Ferrero}},\ and\ \bibinfo {author} {\bibfnamefont {T.}~\bibnamefont {Ayral}},\ }\href@noop {} {\bibfield  {journal} {\bibinfo  {journal} {arXiv:2412.13711}\ } (\bibinfo {year} {2024})}\BibitemShut {NoStop}%
\bibitem [{\citenamefont {S{\o}rensen}\ and\ \citenamefont {M{\o}lmer}(2000)}]{sorensen2000entanglement}%
  \BibitemOpen
  \bibfield  {author} {\bibinfo {author} {\bibfnamefont {A.}~\bibnamefont {S{\o}rensen}}\ and\ \bibinfo {author} {\bibfnamefont {K.}~\bibnamefont {M{\o}lmer}},\ }\href@noop {} {\bibfield  {journal} {\bibinfo  {journal} {Phys. Rev. A}\ }\textbf {\bibinfo {volume} {62}},\ \bibinfo {pages} {022311} (\bibinfo {year} {2000})}\BibitemShut {NoStop}%
\bibitem [{\citenamefont {M{\o}lmer}\ and\ \citenamefont {S{\o}rensen}(1999)}]{molmer1999multiparticle}%
  \BibitemOpen
  \bibfield  {author} {\bibinfo {author} {\bibfnamefont {K.}~\bibnamefont {M{\o}lmer}}\ and\ \bibinfo {author} {\bibfnamefont {A.}~\bibnamefont {S{\o}rensen}},\ }\href@noop {} {\bibfield  {journal} {\bibinfo  {journal} {Phys. Rev. Lett.}\ }\textbf {\bibinfo {volume} {82}},\ \bibinfo {pages} {1835} (\bibinfo {year} {1999})}\BibitemShut {NoStop}%
\bibitem [{\citenamefont {Ma}\ \emph {et~al.}(2022)\citenamefont {Ma}, \citenamefont {Pace},\ and\ \citenamefont {Kim}}]{ma2022unifying}%
  \BibitemOpen
  \bibfield  {author} {\bibinfo {author} {\bibfnamefont {Y.}~\bibnamefont {Ma}}, \bibinfo {author} {\bibfnamefont {M.~C.}\ \bibnamefont {Pace}},\ and\ \bibinfo {author} {\bibfnamefont {M. S.}~\bibnamefont {Kim}},\ }\href@noop {} {\bibfield  {journal} {\bibinfo  {journal} {Phys. Rev. A}\ }\textbf {\bibinfo {volume} {106}},\ \bibinfo {pages} {012605} (\bibinfo {year} {2022})}\BibitemShut {NoStop}%
\bibitem [{\citenamefont {Wang}\ \emph {et~al.}(2020)\citenamefont {Wang}, \citenamefont {Crain}, \citenamefont {Fang}, \citenamefont {Zhang}, \citenamefont {Huang}, \citenamefont {Liang}, \citenamefont {Leung}, \citenamefont {Brown},\ and\ \citenamefont {Kim}}]{wang2020high}%
  \BibitemOpen
  \bibfield  {author} {\bibinfo {author} {\bibfnamefont {Y.}~\bibnamefont {Wang}}, \bibinfo {author} {\bibfnamefont {S.}~\bibnamefont {Crain}}, \bibinfo {author} {\bibfnamefont {C.}~\bibnamefont {Fang}}, \bibinfo {author} {\bibfnamefont {B.}~\bibnamefont {Zhang}}, \bibinfo {author} {\bibfnamefont {S.}~\bibnamefont {Huang}}, \bibinfo {author} {\bibfnamefont {Q.}~\bibnamefont {Liang}}, \bibinfo {author} {\bibfnamefont {P.~H.}\ \bibnamefont {Leung}}, \bibinfo {author} {\bibfnamefont {K.~R.}\ \bibnamefont {Brown}},\ and\ \bibinfo {author} {\bibfnamefont {J.}~\bibnamefont {Kim}},\ }\href@noop {} {\bibfield  {journal} {\bibinfo  {journal} {Phys. Rev. Lett.}\ }\textbf {\bibinfo {volume} {125}},\ \bibinfo {pages} {150505} (\bibinfo {year} {2020})}\BibitemShut {NoStop}%
\bibitem [{\citenamefont {Fang}\ \emph {et~al.}(2022)\citenamefont {Fang}, \citenamefont {Wang}, \citenamefont {Huang}, \citenamefont {Brown},\ and\ \citenamefont {Kim}}]{fang2022crosstalk}%
  \BibitemOpen
  \bibfield  {author} {\bibinfo {author} {\bibfnamefont {C.}~\bibnamefont {Fang}}, \bibinfo {author} {\bibfnamefont {Y.}~\bibnamefont {Wang}}, \bibinfo {author} {\bibfnamefont {S.}~\bibnamefont {Huang}}, \bibinfo {author} {\bibfnamefont {K.~R.}\ \bibnamefont {Brown}},\ and\ \bibinfo {author} {\bibfnamefont {J.}~\bibnamefont {Kim}},\ }\href@noop {} {\bibfield  {journal} {\bibinfo  {journal} {Phys. Rev. Lett.}\ }\textbf {\bibinfo {volume} {129}},\ \bibinfo {pages} {240504} (\bibinfo {year} {2022})}\BibitemShut {NoStop}%
\bibitem [{\citenamefont {Kang}\ \emph {et~al.}(2023)\citenamefont {Kang}, \citenamefont {Wang}, \citenamefont {Fang}, \citenamefont {Zhang}, \citenamefont {Khosravani}, \citenamefont {Kim},\ and\ \citenamefont {Brown}}]{kang2023designing}%
  \BibitemOpen
  \bibfield  {author} {\bibinfo {author} {\bibfnamefont {M.}~\bibnamefont {Kang}}, \bibinfo {author} {\bibfnamefont {Y.}~\bibnamefont {Wang}}, \bibinfo {author} {\bibfnamefont {C.}~\bibnamefont {Fang}}, \bibinfo {author} {\bibfnamefont {B.}~\bibnamefont {Zhang}}, \bibinfo {author} {\bibfnamefont {O.}~\bibnamefont {Khosravani}}, \bibinfo {author} {\bibfnamefont {J.}~\bibnamefont {Kim}},\ and\ \bibinfo {author} {\bibfnamefont {K.~R.}\ \bibnamefont {Brown}},\ }\href@noop {} {\bibfield  {journal} {\bibinfo  {journal} {Phys. Rev. Applied}\ }\textbf {\bibinfo {volume} {19}},\ \bibinfo {pages} {014014} (\bibinfo {year} {2023})}\BibitemShut {NoStop}%
\bibitem [{\citenamefont {Lubinski}\ \emph {et~al.}(2023)\citenamefont {Lubinski}, \citenamefont {Johri}, \citenamefont {Varosy}, \citenamefont {Coleman}, \citenamefont {Zhao}, \citenamefont {Necaise}, \citenamefont {Baldwin}, \citenamefont {Mayer},\ and\ \citenamefont {Proctor}}]{lubinski2023application}%
  \BibitemOpen
  \bibfield  {author} {\bibinfo {author} {\bibfnamefont {T.}~\bibnamefont {Lubinski}}, \bibinfo {author} {\bibfnamefont {S.}~\bibnamefont {Johri}}, \bibinfo {author} {\bibfnamefont {P.}~\bibnamefont {Varosy}}, \bibinfo {author} {\bibfnamefont {J.}~\bibnamefont {Coleman}}, \bibinfo {author} {\bibfnamefont {L.}~\bibnamefont {Zhao}}, \bibinfo {author} {\bibfnamefont {J.}~\bibnamefont {Necaise}}, \bibinfo {author} {\bibfnamefont {C.~H.}\ \bibnamefont {Baldwin}}, \bibinfo {author} {\bibfnamefont {K.}~\bibnamefont {Mayer}},\ and\ \bibinfo {author} {\bibfnamefont {T.}~\bibnamefont {Proctor}},\ }\href@noop {} {\bibfield  {journal} {\bibinfo  {journal} {IEEE Transactions on Quantum Engineering}\ }\textbf {\bibinfo {volume} {4}},\ \bibinfo {pages} {1} (\bibinfo {year} {2023})}\BibitemShut {NoStop}%
\bibitem [{\citenamefont {Sun}\ \emph {et~al.}(2024)\citenamefont {Sun}, \citenamefont {Kang}, \citenamefont {Nuomin}, \citenamefont {Schwartz}, \citenamefont {Beratan}, \citenamefont {Brown},\ and\ \citenamefont {Kim}}]{sun2024quantum}%
  \BibitemOpen
  \bibfield  {author} {\bibinfo {author} {\bibfnamefont {K.}~\bibnamefont {Sun}}, \bibinfo {author} {\bibfnamefont {M.}~\bibnamefont {Kang}}, \bibinfo {author} {\bibfnamefont {H.}~\bibnamefont {Nuomin}}, \bibinfo {author} {\bibfnamefont {G.}~\bibnamefont {Schwartz}}, \bibinfo {author} {\bibfnamefont {D.~N.}\ \bibnamefont {Beratan}}, \bibinfo {author} {\bibfnamefont {K.~R.}\ \bibnamefont {Brown}},\ and\ \bibinfo {author} {\bibfnamefont {J.}~\bibnamefont {Kim}},\ }\href@noop {} {\bibfield  {journal} {\bibinfo  {journal} {arXiv:2405.14624}\ } (\bibinfo {year} {2024})}\BibitemShut {NoStop}%
\bibitem [{\citenamefont {Maslov}(2017)}]{maslov2017basic}%
  \BibitemOpen
  \bibfield  {author} {\bibinfo {author} {\bibfnamefont {D.}~\bibnamefont {Maslov}},\ }\href@noop {} {\bibfield  {journal} {\bibinfo  {journal} {New J. Phys.}\ }\textbf {\bibinfo {volume} {19}},\ \bibinfo {pages} {023035} (\bibinfo {year} {2017})}\BibitemShut {NoStop}%
\bibitem [{nat()}]{native_gates}%
  \BibitemOpen
  \href@noop {} {\bibinfo {title} {Getting started with native gates}},\ \bibinfo {howpublished} {\url{https://ionq.com/docs/getting-started-with-native-gates}},\ \bibinfo {note} {accessed: 2024-05-21}\BibitemShut {NoStop}%
\bibitem [{sup()}]{supp}%
  \BibitemOpen
  \href@noop {} {\bibinfo {title} {See the supplemental information.}}\BibitemShut {Stop}%
\bibitem [{\citenamefont {Monroe}\ \emph {et~al.}(2021)\citenamefont {Monroe}, \citenamefont {Campbell}, \citenamefont {Duan}, \citenamefont {Gong}, \citenamefont {Gorshkov}, \citenamefont {Hess}, \citenamefont {Islam}, \citenamefont {Kim}, \citenamefont {Linke}, \citenamefont {Pagano} \emph {et~al.}}]{monroe2021programmable}%
  \BibitemOpen
  \bibfield  {author} {\bibinfo {author} {\bibfnamefont {C.}~\bibnamefont {Monroe}}, \bibinfo {author} {\bibfnamefont {W.~C.}\ \bibnamefont {Campbell}}, \bibinfo {author} {\bibfnamefont {L.-M.}\ \bibnamefont {Duan}}, \bibinfo {author} {\bibfnamefont {Z.-X.}\ \bibnamefont {Gong}}, \bibinfo {author} {\bibfnamefont {A.~V.}\ \bibnamefont {Gorshkov}}, \bibinfo {author} {\bibfnamefont {P.~W.}\ \bibnamefont {Hess}}, \bibinfo {author} {\bibfnamefont {R.}~\bibnamefont {Islam}}, \bibinfo {author} {\bibfnamefont {K.}~\bibnamefont {Kim}}, \bibinfo {author} {\bibfnamefont {N.~M.}\ \bibnamefont {Linke}}, \bibinfo {author} {\bibfnamefont {G.}~\bibnamefont {Pagano}}, \emph {et~al.},\ }\href@noop {} {\bibfield  {journal} {\bibinfo  {journal} {Rev. Mod. Phys.}\ }\textbf {\bibinfo {volume} {93}},\ \bibinfo {pages} {025001} (\bibinfo {year} {2021})}\BibitemShut {NoStop}%
\bibitem [{\citenamefont {Foss-Feig}\ \emph {et~al.}(2024)\citenamefont {Foss-Feig}, \citenamefont {Pagano}, \citenamefont {Potter},\ and\ \citenamefont {Yao}}]{foss2024progress}%
  \BibitemOpen
  \bibfield  {author} {\bibinfo {author} {\bibfnamefont {M.}~\bibnamefont {Foss-Feig}}, \bibinfo {author} {\bibfnamefont {G.}~\bibnamefont {Pagano}}, \bibinfo {author} {\bibfnamefont {A.~C.}\ \bibnamefont {Potter}},\ and\ \bibinfo {author} {\bibfnamefont {N.~Y.}\ \bibnamefont {Yao}},\ }\href@noop {} {\bibfield  {journal} {\bibinfo  {journal} {arXiv:2409.02990}\ } (\bibinfo {year} {2024})}\BibitemShut {NoStop}%
\bibitem [{\citenamefont {Nam}\ \emph {et~al.}(2020)\citenamefont {Nam}, \citenamefont {Chen}, \citenamefont {Pisenti}, \citenamefont {Wright}, \citenamefont {Delaney}, \citenamefont {Maslov}, \citenamefont {Brown}, \citenamefont {Allen}, \citenamefont {Amini}, \citenamefont {Apisdorf} \emph {et~al.}}]{nam2020ground}%
  \BibitemOpen
  \bibfield  {author} {\bibinfo {author} {\bibfnamefont {Y.}~\bibnamefont {Nam}}, \bibinfo {author} {\bibfnamefont {J.-S.}\ \bibnamefont {Chen}}, \bibinfo {author} {\bibfnamefont {N.~C.}\ \bibnamefont {Pisenti}}, \bibinfo {author} {\bibfnamefont {K.}~\bibnamefont {Wright}}, \bibinfo {author} {\bibfnamefont {C.}~\bibnamefont {Delaney}}, \bibinfo {author} {\bibfnamefont {D.}~\bibnamefont {Maslov}}, \bibinfo {author} {\bibfnamefont {K.~R.}\ \bibnamefont {Brown}}, \bibinfo {author} {\bibfnamefont {S.}~\bibnamefont {Allen}}, \bibinfo {author} {\bibfnamefont {J.~M.}\ \bibnamefont {Amini}}, \bibinfo {author} {\bibfnamefont {J.}~\bibnamefont {Apisdorf}}, \emph {et~al.},\ }\href@noop {} {\bibfield  {journal} {\bibinfo  {journal} {npj Quantum Inf.}\ }\textbf {\bibinfo {volume} {6}},\ \bibinfo {pages} {33} (\bibinfo {year} {2020})}\BibitemShut {NoStop}%
\bibitem [{\citenamefont {Wright}\ \emph {et~al.}(2019)\citenamefont {Wright}, \citenamefont {Beck}, \citenamefont {Debnath}, \citenamefont {Amini}, \citenamefont {Nam}, \citenamefont {Grzesiak}, \citenamefont {Chen}, \citenamefont {Pisenti}, \citenamefont {Chmielewski}, \citenamefont {Collins} \emph {et~al.}}]{wright2019benchmarking}%
  \BibitemOpen
  \bibfield  {author} {\bibinfo {author} {\bibfnamefont {K.}~\bibnamefont {Wright}}, \bibinfo {author} {\bibfnamefont {K.~M.}\ \bibnamefont {Beck}}, \bibinfo {author} {\bibfnamefont {S.}~\bibnamefont {Debnath}}, \bibinfo {author} {\bibfnamefont {J.}~\bibnamefont {Amini}}, \bibinfo {author} {\bibfnamefont {Y.}~\bibnamefont {Nam}}, \bibinfo {author} {\bibfnamefont {N.}~\bibnamefont {Grzesiak}}, \bibinfo {author} {\bibfnamefont {J.-S.}\ \bibnamefont {Chen}}, \bibinfo {author} {\bibfnamefont {N.}~\bibnamefont {Pisenti}}, \bibinfo {author} {\bibfnamefont {M.}~\bibnamefont {Chmielewski}}, \bibinfo {author} {\bibfnamefont {C.}~\bibnamefont {Collins}}, \emph {et~al.},\ }\href@noop {} {\bibfield  {journal} {\bibinfo  {journal} {Nat. Commun.}\ }\textbf {\bibinfo {volume} {10}},\ \bibinfo {pages} {5464} (\bibinfo {year} {2019})}\BibitemShut {NoStop}%
\bibitem [{bra()}]{braketConvention}%
  \BibitemOpen
  \href@noop {} {\bibinfo {title} {braket circuits gate module}},\ \bibinfo {howpublished} {\url{https://amazon-braket-sdk-python.readthedocs.io/en/latest/_apidoc/braket.circuits.html}},\ \bibinfo {note} {accessed: 2024-05-21}\BibitemShut {NoStop}%
\bibitem [{\citenamefont {Wilson}(1927)}]{wilson_probable_1927}%
  \BibitemOpen
  \bibfield  {author} {\bibinfo {author} {\bibfnamefont {E.~B.}\ \bibnamefont {Wilson}},\ }\href {https://doi.org/10.1080/01621459.1927.10502953} {\bibfield  {journal} {\bibinfo  {journal} {Journal of the American Statistical Association}\ }\textbf {\bibinfo {volume} {22}},\ \bibinfo {pages} {209} (\bibinfo {year} {1927})}\BibitemShut {NoStop}%
\end{thebibliography}
\end{document}


\title{Supplemental Material: Reshaping quantum device noise via quantum error correction}

\author{Yue Ma$^1$}

\author{Michael Hanks$^1$}

\author{Evdokia Gneusheva$^1$}

\author{M. S. Kim$^1$}

\affiliation{$^1$Blackett Laboratory, Imperial College London, SW7 2AZ, United Kingdom\\
}

\begin{abstract}
In this Supplemental Material, we first analyze circuit-agnostic models for our modified repetition code. Next, we present the analytical and numerical details of deriving the noise of M{\o}lmer-S{\o}rensen gates under dephasing of the oscillation mode. We then explain the technical details of implementing the quantum circuits, Fig.~2 in the main text, on the IonQ Aria-1 quantum device. We next give a detailed description of conventions and gate decompositions that we have used. Afterwards we provide the full compiled circuits that we implement on the IonQ Aria-1 quantum device. Then we present our results of noisy simulations of the compiled quantum circuits. Finally we demonstrate the reshaping of noise with the three-qubit bit-flip repetition code on a superconducting qubit quantum device, the IQM Garnet, and characterize its native Controlled-Z (CZ) gate.

\end{abstract}


\maketitle

\section{Inserting a general single-qubit gate to the bit-flip repetition code: the circuit-agnostic model}\label{sec:U1}

Here we present the analytical result following the circuit-agnostic model of our modified bit-flip repetition code. The circuit-agnostic model is defined as following. The state $|\psi\rangle=\alpha|0\rangle+\beta|1\rangle$ is perfectly encoded into the state $|\bar{\psi}\rangle=\alpha|000\rangle+\beta|111\rangle$. Then, each of the three physical qubits has probability $p$ of being subject to the bit-flip error describe by Pauli-$X$. Then, we modify the code by adding a unitary gate on the first qubit. Finally we perform perfect syndrome measurements and corrections to map the state back to the logical space, resulting in $\bar{\rho}_f$.

Any unitary operator applied to one qubit can be parameterised by three angles:~\cite{nemoto2000generalized}
\begin{equation}\label{eq:U}
U(\theta,\phi,\delta)=
    \begin{pmatrix}
        \exp(i\phi)\cos(\theta) & -\exp(-i\delta)\sin(\theta)\\
        \exp(i\delta)\sin(\theta) & \exp(-i\phi)\cos(\theta)
    \end{pmatrix},
\end{equation}
where $0\leq\theta\leq\pi/2$, $0\leq\phi\leq2\pi$, $0\leq\delta\leq2\pi$. After syndrome measurements and corrections, the final state is derived to be
\begin{align}
    \bar{\rho}_f&=(1+p)(1-p)^2\cos^2(\theta)R_1|\bar{\psi}\rangle\langle\bar{\psi}|R_1^{\dagger}\nonumber\\
    &+(2-p)(1-p)p\sin^2(\theta)R_2|\bar{\psi}\rangle\langle\bar{\psi}|R_2^{\dagger}\nonumber\\
    &+(2-p)p^2\cos^2(\theta)R_3|\bar{\psi}\rangle\langle\bar{\psi}|R_3^{\dagger}\nonumber\\
    &+p(1+p)(1-p)\sin^2(\theta)R_4|\bar{\psi}\rangle\langle\bar{\psi}|R_4^{\dagger}\nonumber\\
    &+(1-p)^3\sin^2(\theta)R_5|\bar{\psi}\rangle\langle\bar{\psi}|R_5^{\dagger}\nonumber\\
    &+p^2(1-p)\cos^2(\theta)R_6|\bar{\psi}\rangle\langle\bar{\psi}|R_6^{\dagger}\nonumber\\
    &+p^3\sin^2(\theta)R_7|\bar{\psi}\rangle\langle\bar{\psi}|R_7^{\dagger}\nonumber\\
    &+p(1-p)^2\cos^2(\theta)R_8|\bar{\psi}\rangle\langle\bar{\psi}|R_8^{\dagger},
\end{align}
with
\begin{equation*}
R_1=
    \begin{pmatrix}
        \exp(i\phi) & 0\\
        0 & \exp(-i\phi)
    \end{pmatrix},
R_2=
    \begin{pmatrix}
        0 & -\exp(-i\delta)\\
        \exp(i\delta) & 0
    \end{pmatrix},
\end{equation*}
\begin{equation*}
R_3=
    \begin{pmatrix}
        0 & \exp(i\phi)\\
        \exp(-i\phi) & 0
    \end{pmatrix},
R_4=
    \begin{pmatrix}
        -\exp(-i\delta) & 0\\
        0 & \exp(i\delta)
    \end{pmatrix},    
\end{equation*}
\begin{equation*}
R_5=
    \begin{pmatrix}
        \exp(i\delta) & 0\\
        0 & -\exp(-i\delta)
    \end{pmatrix},
R_6=
    \begin{pmatrix}
        0 & \exp(-i\phi)\\
        \exp(i\phi) & 0
    \end{pmatrix},
\end{equation*}
\begin{equation*}
R_7=
    \begin{pmatrix}
        0 & \exp(i\delta)\\
        -\exp(-i\delta) & 0
    \end{pmatrix},
R_8=
    \begin{pmatrix}
        \exp(-i\phi) & 0\\
        0 & \exp(i\phi)
    \end{pmatrix}.    
\end{equation*}

For the special case of choosing the unitary gate on qubit 1 as the $R_y(\theta_0)$ gate as defined in the main text, we can get simple expressions: The resulting output state is expressed as the initial logical state $|\bar{\psi}\rangle$ going through a quantum channel described by the following Kraus operators,
\begin{align}\label{eq:ANAstate}
    K_I&=(1-p)\sqrt{1+2p}\cos(\theta_0/2)I_L,\nonumber\\
    K_X&=\sqrt{3p^2-2p^3}\cos(\theta_0/2)X_L,\nonumber\\
    K_Y&=\sqrt{2p-3p^2+2p^3}\sin(\theta_0/2)Y_L,\nonumber\\
    K_Z&=\sqrt{1-2p+3p^2-2p^3}\sin(\theta_0/2)Z_L,
\end{align}
where $I_L=I_1I_2I_3$ is the logical identity operator, $X_L=X_1X_2X_3$, $Z_L = Z_1$ and $Y_L=-iZ_LX_L$ are the logical Pauli-$X$, $Z$ and $Y$ operators, respectively. Compared with the physical bit-flip channel, this channel has reduced $X$ error, but has $Y$ and $Z$ errors that are not present on the physical qubit level. The correlations between the weights of the errors are also nontrivial. If we choose $\theta_0=0$, that is, no $R_y$ gate is acted on qubit 1, the channel expressed by Eq.~\eqref{eq:ANAstate} is consistent with the success probability $p_{\mathrm{success}}=(1-p)^3+3(1-p)^2p$ of the code. As the value of $\theta_0$ is increased, logical $Y$ and $Z$ operators start to contribute to the channel. If we take $\theta_0=\pi$, the channel is entirely made of the stochastic mixture of logical $Y$ and $Z$ operators. These nontrivial relations are fundamentally different from the partial probabilistic error cancellation scheme in Ref.~\cite{guimaraes2023noise}, where classical post-processing modifies the weight of each Pauli operator individually. The correlation in Eq.~\eqref{eq:ANAstate} implies that our method of reshaping the native noise via quantum error correction codes has the potential of defining novel nontrivial hierarchies for what target quantum channels are easier to be simulated, even when the original noise channels do not have particular structures (e.g. Pauli noise). It is also worth emphasizing that the tunability of the channel upon changing $\theta_0$ relies on the non-zero bit-flip rate $p$: We are exploiting the noise as a useful feature to unveil the abundant structures of the reshaped channel.

\begin{figure}[t!]
    \centering
    \includegraphics[width=0.48\textwidth]{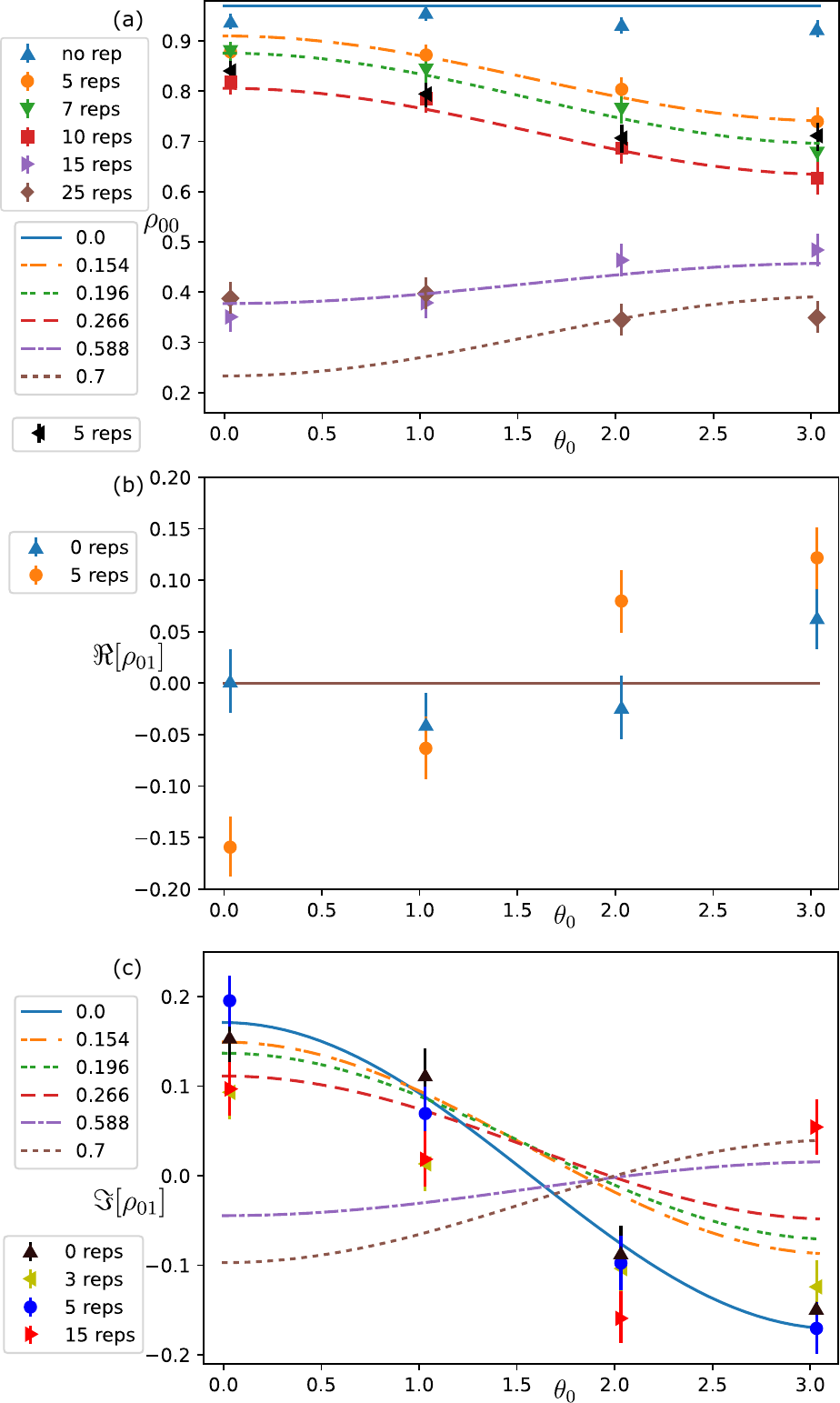}
    \caption{\label{figSUPP:data_rho00}
    Measurement results obtained from the IonQ Aria-1 quantum device (data points), in comparison with the circuit-agnostic model Eq.~\eqref{eq:ANAstate} (lines). The data points are the same as described in the main text. For the lines, each bit-flip rate $p$ is an integer multiple of the M{\o}lmer-S{\o}rensen gate infidelity, $0.014$ at the time of measurements. (a) The diagonal element of the output logical density matrix, ${\rho_{00}}$. (b\{c\}) The real \{imaginary\} part of the off-diagonal element $\rho_{01}$.
    }
\end{figure}

In Fig.~\ref{figSUPP:data_rho00}, we demonstrate that the circuit-agnostic analytical model matches reasonably well with the data we got from the IonQ Aria-1 quantum device.

\section{Details of noisy M{\o}lmer-S{\o}rensen gates} 
The implementation of M{\o}lmer-S{\o}rensen gates~\cite{molmer1999multiparticle,sorensen2000entanglement} in trapped-ion systems involve two ions whose intrinsic energy levels are considered as qubits, and their common vibrational mode which is modelled as a mechanical oscillator whose virtual excitation by the laser field is used to entangle the intrinsic states of the two ions. The Hamiltonian is generally referred to as~\cite{sorensen2000entanglement}
\begin{equation}\label{eq:Hamiltonian}
    H=f(t)J_yx+g(t)J_yp,
\end{equation}
where $J_y=(\sigma_{y1}+\sigma_{y2})/2$, $\sigma_{yi}$ are Pauli operators, $f(t)=-\sqrt{2}\eta\Omega\cos(\nu-\delta)t$, $g(t)=-\sqrt{2}\eta\Omega\sin(\nu-\delta)t$, where $\eta$ is the Lamb-Dicke parameter, $\Omega$ is the Rabi frequency, $\nu$ is the mechanical frequency,  $\delta$ is the laser detuning, $x$ and $p$ are the dimensionless position and momentum operators, respectively. 
We want to derive the quantum channel that connects the final state subject to noise with the target state. We therefore work in the interaction picture with respect to the unitary evolution governed by Eq.~\eqref{eq:Hamiltonian}~\cite{sorensen2000entanglement}. In this way, the dynamics in the interaction picture directly correspond to our desired quantum channel. 

The dominant source of noise for trapped ion systems is the motional dephasing~\cite{sun2024quantum}. In the Lindblad form, this corresponds to the dissipator $C=\sqrt{\Gamma_P}a^{\dagger}a$. After moving into the interaction picture, the dissipator becomes
\begin{equation}
    \tilde{C}=\sqrt{\Gamma_P}(a^{\dagger}+J_y\frac{G(t)+iF(t)}{\sqrt{2}})(a+J_y\frac{G(t)-iF(t)}{\sqrt{2}}),
\end{equation}
where $G(t)=\int_0^tdt'g(t')$, $F(t)=\int_0^tdt'f(t')$, and the dynamics become
\begin{equation}\label{eq:fullDyn}
    \frac{d}{dt}\tilde{\rho}(t)=-\frac{1}{2}\tilde{C}^{\dagger}\tilde{C}\tilde{\rho}-\frac{1}{2}\tilde{\rho}\tilde{C}^{\dagger}\tilde{C}+\tilde{C}\tilde{\rho}\tilde{C}^{\dagger}.
\end{equation}
We approximate that throughout the evolution, in this interaction picture, the state of the two qubits remains separable from the state of the oscillator, which stays in the initial thermal state. Specifically, we assume $\tilde{\rho}(t)\approx\tilde{\rho}_q(t)\otimes\gamma_{\nu}$, where $\tilde{\rho}_q(t)$ is the joint state of the two qubits, and $\gamma_{\nu}$ is the thermal state of the oscillator with mean phonon number $n_{\mathrm{th}}$. Under this approximation and taking the partial trace over the oscillator mode, Eq.~\eqref{eq:fullDyn} becomes the dynamical equation for only the state of the qubits,
\begin{align}
    &\frac{d}{dt}\tilde{\rho}_q(t)=\nonumber\\
    &(G^2(t)+F^2(t))(n_{\mathrm{th}}+\frac{1}{2})\Gamma_P(-\frac{1}{2}J_y^2\tilde{\rho}_q-\frac{1}{2}\tilde{\rho}_qJ_y^2+J_y\tilde{\rho}_qJ_y)\nonumber\\
    &+\frac{1}{4}(G^2(t)+F^2(t))^2\Gamma_P(-\frac{1}{2}J_y^4\tilde{\rho}_q-\frac{1}{2}\tilde{\rho}_qJ_y^4+J_y^2\tilde{\rho}_qJ_y^2).
\end{align}
The equation can be straightforwardly solved in an element-wise manner by expanding all operators in the eigenbasis of $J_y$~\cite{sorensen2000entanglement}. We are interested in interaction times when the state of the qubits disentangles with the state of the oscillator in the original picture, and when the maximal entanglement between the two qubits is generated (up to local unitary operators that rotate $Y$ into $X$, this will be referred to as XX gates later). These correspond to the constraints between parameters~\cite{sorensen2000entanglement}, $\tau=2\pi K/(\nu-\delta)$ and $\eta\Omega/(\nu-\delta)=1/2\sqrt{K}$, where $K$ is an integer referring to how many phase-space loops of the oscillator have been traversed~\cite{ma2022unifying}. For simplicity, we define two parameters
\begin{align}
    r_1&=\frac{1}{4}(2n_{\mathrm{th}}+1)\Gamma_P\frac{\tau}{K},\\
    r_2&=\frac{1}{8}\Gamma_P\frac{3\tau}{2K^2}.
\end{align}
The quantum channel that maps the initial state $\tilde{\rho}_q(0)$ to the final state $\tilde{\rho}_q(\tau)$, is found to have the following canonical Kraus operators ($\mathrm{Tr}(K_i^{\dagger}K_j)=0$ for $i\neq j$)
\begin{align}\label{eq:Kraus}
    K_1&=a_1I+a_3J_y^2,\nonumber\\
    K_2&=\sqrt{\frac{1}{2}(1-e^{-4r_1})}J_y,\nonumber\\
    K_3&=a_2I+a_4J_y^2,
\end{align}
where $a_1$, $a_2$, $a_3$ and $a_4$ are solutions of the coupled equations
\begin{align}\label{eq:coupledEqs}
    &a_1^2+a_2^2=1,\nonumber\\
    &a_1a_3+a_2a_4=e^{-r_1-r_2}-1,\nonumber\\
    &2a_1a_3+a_3^2+2a_2a_4+a_4^2=\frac{1}{2}(e^{-4r_1}-1),\nonumber\\
    &2a_1a_2+a_3a_4+a_1a_4+a_2a_3=0.
\end{align}
Note that, the canonical Kraus representation of a quantum channel is unique. The non-uniquenesses in the solution of Eq.~\eqref{eq:coupledEqs} are that we can simulatneously swap $a_1$ with $a_2$ and $a_3$ with $a_4$, and that there are global phases for each Kraus operator in Eq.~\eqref{eq:Kraus}. Considering that practical M{\o}lmer-S{\o}rensen gates operate in the weak coupling regime~\cite{monroe2021programmable,foss2024progress,nam2020ground,wright2019benchmarking}, i.e, small phase-space loops are traversed many times, one canonical Kraus operator has a dominant identity component. We fix this to be $K_1$. We find that $a_1\approx1$ has closed form solution, after removing the global phases by requiring $a_1>0$ and $a_2>0$,
\begin{widetext}
   \begin{equation}\label{eq:a1}
    a_1=\sqrt{\frac{e^{-r_2}(4e^{2r_1}-e^{2r_2}+e^{4r_1+2r_2}+e^{r_2}\sqrt{16e^{6r_1}+e^{2r_2}(-1+e^{4r_1})^2})}{2\sqrt{16e^{6r_1}+e^{2r_2}(-1+e^{4r_1})^2}}}
\end{equation} 
\end{widetext}
The solution for $a_2$, $a_3$ and $a_4$ will be found once Eq.~\eqref{eq:a1} is inserted into Eq.~\eqref{eq:coupledEqs}. It should be pointed out that for the weak coupling regime, $r_2\ll r_1$ and $r_2\approx 0$. This implies the coefficient of $K_2$, which is defined as $k_2$ in the main text, will uniquely determine what $a_1,a_2,a_3,a_4$ are through $r_1$. 

As a practical example, we consider the weak-coupling parameters given in Ref.~\cite{sorensen2000entanglement}, $\eta=0.1$, $\Omega=0.1\nu$, $K=25$ which comes from taking $\delta=0.9\nu$. For $\Gamma_P=0.02\nu$ and $n_{\mathrm{th}}=0.05$, we find $r_1=0.3456$ and $r_2=0.0094$. These lead to the solution $a_1=0.9837$, $a_2=0.1797$, $a_3=-0.2282$, $a_4=-0.4136$. In particular, $K_2=0.612J_y$ is the leading noise source, while $K_1$ is close to the identity operator. Importantly, $K_2$ represents the sum of bit-flip errors for each of the two qubits. Note that in theoretical derivations~\cite{sorensen2000entanglement}, the target gate is in the YY form, while for experimental trapped ion systems such as IonQ, the target gate is in the XX form. This implies that $K_2$ in the experimental XX gate setup will be proportional to $\sigma_{x1}+\sigma_{x2}$, which is indeed the sum of bit-flip errors. We will show that this sum will be discretized by the syndrome measurements of the bit-flip repetition code, therefore the noisy M{\o}lmer-S{\o}rensen gates are directly compatible with the error assumptions of the codes.

We also numerically simulated the exact dynamics following the Hamiltonian Eq.~\eqref{eq:Hamiltonian} and the dephasing dissipator $C$. Using the diagonalization of the Choi matrix, we find that the structure of the canonical Kraus operators in Eq.~\eqref{eq:Kraus} remains the same, while each parameter is slightly different. The results are shown in Fig.~1 in the main text. The numerical results match reasonably well with the analytical ones. In particular, $K_2$ represents the dominant error term. Here we also list the relevant numbers. For the numerical results shown in the main text Fig.~3, the coefficients of the channel Kraus operators are numerically calculated based on assuming $\Gamma_P=0.0025\nu$, i.e., the left-most set of points in the main text Fig.~1. The result is: $a_1=0.999+0.00024i$, $a_2=0.03647 + 0.00063i$, $a_3=-0.028 - 0.00048i$, $a_4=-0.074 - 0.00126i$, $k_2=0.24$.


For comparison, we also derive the quantum channel for the heating of the oscillator. The approximate analytical solution of the master equation is already derived in Ref.~\cite{sorensen2000entanglement}, however, there the fidelity was calculated rather than the quantum channel. We find that, the canonical Kraus form of the quantum channel is also represented by Eq.~\eqref{eq:Kraus}. The coefficients correspond to directly replacing $r_1\rightarrow r$, $r_2\rightarrow 0$, $\Gamma_P\rightarrow\Gamma$, and $n_{\mathrm{th}}$ becoming the mean bath phonon number for the relaxation operators~\cite{sorensen2000entanglement}. Exact numerical simulation also leads to the same structure as Eq.~\eqref{eq:Kraus}. We can therefore deduce that based on the state of the two qubits alone, we cannot distinguish whether the oscillator mode is subject to dephasing noise or heating noise.



\section{Details of implementing noise reshaping on IonQ Aria-1} 

The noisiest quantum operations on the IonQ Aria-1 quantum device are the two-qubit gates~\cite{IonQ_Fidelity}. Two-qubit gates are essential for generating entanglement, in particular, for encoding into the code space and syndrome measurements in our modified repetition code. IonQ uses the M{\o}lmer-S{\o}rensen gate~\cite{molmer1999multiparticle} as the native two-qubit gate~\cite{native_gates}. Specifically, it exploits the $X$-$X$ type interaction, $\exp[-i(\pi/4)X_1X_2]$ for entangling qubits $1$ and $2$, which is commonly referred to as the XX gate~\cite{trout2018simulating}. This gate is equivalent to the more conventionally used CNOT gate upto local rotations (see Sec.~\ref{sec:convention}, where all the conventions are also defined).

The quantum circuit for implementing our modified three-qubit bit-flip repetition code is the one in Fig.~2(a) in the main text. We have chosen to use four values of $\theta_0$ in the $R_y$ rotation, ${\theta_{0} / \pi \in\{ 0.01 , 0.328 , 0.647 , 0.965 \}}$. The $R_x$ gate corresponds to the initialisation to $\alpha|0\rangle+\beta|1\rangle$, where $\alpha=\cos(\phi/2)$ and $\beta=-i\sin(\phi/2)$. We have chosen $\phi=\pi/9$. The two CNOT gates between the $R_x$ gate and the $R_y$ gate correspond to the encoding process to the state $|\bar{\psi}\rangle$. The four CNOT gates after the $R_y$ gate are for the syndrome measurement and correction process. Qubits 4 and 5 are the ancillae and are measured in the computational basis. The outcomes $(0,0)$, $(0,1)$, $(1,0)$ and $(1,1)$ correspond to the four lines in Table 1 in the main text, respectively. Correction gates should be applied accordingly to bring the logical state back to the code space, which is spanned by $\{|000\rangle,|111\rangle\}$. However, as Amazon Braket does not support mid-circuit measurements, here we directly measure the three physical qubits (labelled 1,2,3) in the computational basis. Post-processing of the classical measurement results after many shots lead to the statistical estimation of the diagonal element of the final logical state $\bar{\rho}_f$, i.e., $\langle000|\bar{\rho}_f|000\rangle\equiv\rho_{00}$. Note that, this scheme including two ancillae can account for some measurement errors, as we only count measurement results that are compatible with the code space basis as valid.

The circuit in Fig.~2(a) in the main text needs to be rewritten into the combination of IonQ native gates, which include GPi1 and GPi2 for single-qubit gates and M\o lmer-S\o rensen gates for two-qubit entangling gates~\cite{native_gates}. We manually compile the circuit in order to keep a clear structure. To be specific, we first decompose the CNOT gate in terms of the $XX$ gate and $R_x$, $R_y$, $R_z$ gates (see Sec.~\ref{sec:convention}). We then convert these gates into the native gates, keeping track of the phase of each qubit so that a virtual $R_z$ gate is applied to each qubit at the end of the circuit~\cite{native_gates}. The virtual $R_z$ gates are not actually implemented at the hardware level as the qubits are only measured in the basis $\{|0\rangle,|1\rangle\}$. However, as they are propagated till the end of the circuit, it is no longer possible to separate the circuit into encoding section and decoding section. Instead, what is clear is that the first two M\o lmer-S\o rensen gates, which are between qubits labelled as $(1,2)$ and as $(1,3)$, encode the quantum information into the logical state, as they are associated with the first two CNOT gates in Fig.~2(a) in the main text. Moreover, they are applied before the gates related to the tunable parameter $\theta_0$ are applied. This gives us a direct connection with the circuit-agnostic analytical model, Eq.~\eqref{eq:ANAstate}, where the stochastic bit-flip errors are applied before the unitary rotation. Indeed, given that the implementation of M\o lmer-S\o rensen gates is much noisier than the GPi1 and GPi2 gates, we add in the tunability of the level of noise in the circuit by repeatedly inserting~\cite{wang2020high,fang2022crosstalk} different numbers of redundant pairs of M\o lmer-S\o rensen gates that correspond to the identity operator in the noiseless case, after each of the two encoding M\o lmer-S\o rensen gates (see Sec.~\ref{sec:compiledCircuit} for the actual circuits we have run on the quantum device).

The diagonal element $\rho_{00}$ itself does not provide a full picture of the logical channel which is reshaped from the native device noise via our modified bit-flip repetition codes. We are going to show that we have also measured the off-diagonal element, $\langle000|\bar{\rho}_f|111\rangle\equiv\rho_{01}$, from the quantum device. However, if we were to directly get $\rho_{01}$ from the quantum circuit Fig.~2(a) in the main text, quantum gates that involve the weighted sum of the logical operators $X_L$ and $Z_L$ need to be applied, which act on the three physical qubits simultaneously and are thus difficult and noisy to implement on the IonQ device. We therefore make use of different quantum circuits here to experimentally extract the complex value of $\rho_{01}$, using only three qubits without ancillae. The circuits are shown in Fig.~2(b) in the main text. It is straightforward to check that this circuit (without the final gate $U_{\rho}$ for extracting the logical density matrix elements via computational basis measurements) is equivalent to the one in Fig.~2(a) in the main text, using ZX calculus~\cite{hanks2020effective}. The value of $\rho_{01}$ (and $\rho_{00}$ for comparison with Fig.~2(a) in the main text) is constructed in the following way.

\noindent\textit{Case 1.} $U_{\rho}$ is chosen as the identity operator. We construct:
\begin{equation}
    \frac{\mathrm{the\ number\ of\ trials\ in\ the\ table\ below}}{\mathrm{the\ total\ number\ of\ trials}}\Rightarrow\rho_{00}
\end{equation}
with the table as
\begin{table}[h]
    \centering
    \begin{tabular}{c|c|c}
        qubit 1 & qubit 2 & qubit 3 \\
        \hline
        0 & 0 & 0 \\
        0 & 0 & 1 \\
        0 & 1 & 0 \\
        1 & 1 & 1
    \end{tabular}.
\end{table}

\noindent\textit{Case 2.} $U_{\rho}$ is chosen as $R_y(\pi/2)R_z(-\pi)$. We construct:
\begin{equation}
    \frac{\mathrm{the\ number\ of\ trials\ in\ the\ table\ below}}{\mathrm{the\ total\ number\ of\ trials}}-\frac{1}{2}\Rightarrow\Re[\rho_{01}]
\end{equation}
with the table as
\begin{table}[h]
    \centering
    \begin{tabular}{c|c|c}
        qubit 1 & qubit 2 & qubit 3 \\
        \hline
        0 & 0 & 0 \\
        0 & 0 & 1 \\
        0 & 1 & 0 \\
        0 & 1 & 1
    \end{tabular}.
\end{table}

\noindent\textit{Case 3.} $U_{\rho}$ is chosen as $R_y(\pi/2)R_z(-\pi/2)$. We construct:
\begin{equation}
    \frac{\mathrm{the\ number\ of\ trials\ in\ the\ table\ below}}{\mathrm{the\ total\ number\ of\ trials}}-\frac{1}{2}\Rightarrow\Im[\rho_{01}]
\end{equation}
with the table as
\begin{table}[h]
    \centering
    \begin{tabular}{c|c|c}
        qubit 1 & qubit 2 & qubit 3 \\
        \hline
        0 & 0 & 0 \\
        0 & 0 & 1 \\
        0 & 1 & 0 \\
        1 & 1 & 1
    \end{tabular}.
\end{table}

\section{Conventions and gate decompositions}\label{sec:convention}

In this section, we first clarify the conventions we have followed for using the IonQ Aria-1 quantum device via Amazon Braket. We then illustrate the gate decompositions, namely, how CNOT gate, XX gate and M{\o}lmer-S{\o}rensen gates with different phases are related.

\begin{figure}[h]
    \centering
    \includegraphics[width=0.48\textwidth]{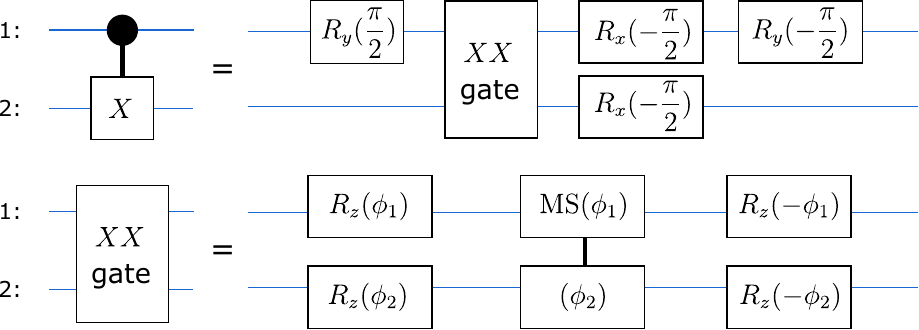}
    \caption{\label{fig:MStoCNOT}
        The relation between the CNOT gate, the XX gate and the M{\o}lmer-S{\o}rensen gates. \textbf{(Top)} The CNOT gate can be decomposed as a combination of single-qubit rotations and the XX gate. \textbf{(Bottom)} The XX gate can be decomposed as the M{\o}lmer-S{\o}rensen gate $\mathrm{MS}(\phi_1,\phi_2)$ sandwiched by single-qubit $R_z$ rotations.
    }
\end{figure}

For expressing quantum operators as matrices, we follow the conventions of Amazon Braket~\cite{braketConvention}. Specifically, the qubit computational basis is given by
\begin{equation}
|0\rangle=
    \begin{pmatrix}
    1\\
    0
    \end{pmatrix},\ \ 
|1\rangle=
    \begin{pmatrix}
    0\\
    1
    \end{pmatrix}.    
\end{equation}
The CNOT gate, where qubit 1 is the control and qubit 2 is the target, is
\begin{align}
    \mathrm{CNOT}&=|0\rangle_1\langle0|\otimes I_2+|1\rangle_1\langle1|\otimes X_2\nonumber\\
    &=
        \begin{pmatrix}
        1 & 0 & 0 & 0\\
        0 & 1 & 0 & 0\\
        0 & 0 & 0 & 1\\
        0 & 0 & 1 & 0
    \end{pmatrix}.
\end{align}
The single-qubit rotations are
\begin{equation}\label{eq:Rx}
R_x(\phi)=
    \begin{pmatrix}
        \cos(\phi/2) & -i\sin(\phi/2)\\
        -i\sin(\phi/2) & \cos(\phi/2)
    \end{pmatrix},
\end{equation}
and
\begin{equation}\label{eq:Rz}
R_z(\phi)=
    \begin{pmatrix}
        \exp(-i\phi/2) & 0\\
        0 & \exp(i\phi/2)
    \end{pmatrix},
\end{equation}
while $R_y$ is already introduced in Eq.~(2) in the main text. M{\o}lmer-S{\o}rensen gates, which are the native two-qubit entangling gates implemented on IonQ Aria-1, include two phase parameters $\phi_1$ and $\phi_2$,
\begin{align}
    &\mathrm{MS}(\phi_1,\phi_2)=\frac{1}{\sqrt{2}}\times\nonumber\\
   & \begin{pmatrix}
        1 & 0 & 0 & -ie^{-i(\phi_1+\phi_2)}\\
        0 & 1 & -ie^{-i(\phi_1-\phi_2)} & 0\\
        0 & -ie^{i(\phi_1-\phi_2)} & 1 & 0\\
        -ie^{i(\phi_1+\phi_2)} & 0 & 0 & 1
    \end{pmatrix}.
\end{align}
Their relation to the CNOT gate via the XX gate is illustrated in Fig.~\ref{fig:MStoCNOT}.


\begin{figure}[h]
    \centering
    \includegraphics[width=0.48\textwidth]{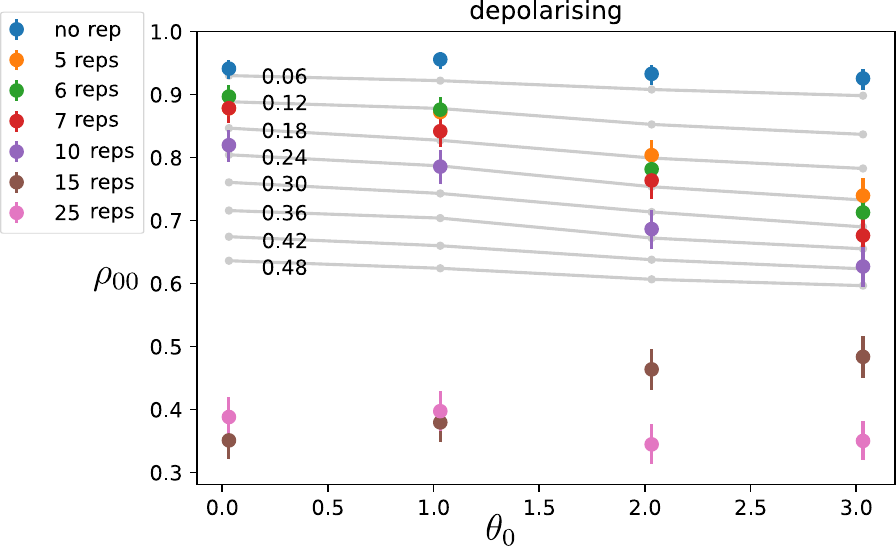}
    \caption{\label{fig:circuitSimDpl}
    Comparing the data from the IonQ Aria-1 device with the results from the noisy simulations of quantum circuits, for the value of $\rho_{00}$, i.e., the diagonal element of the output logical state. The points with error bars correspond to the data, where certain numbers of redundant pairs of M{\o}lmer-S{\o}rensen gates have been added in the quantum circuit with native gates in order to tune the error rate. Each data point corresponds to an average over ${1000}$ shots. Error bars are ${95\%}$ confidence intervals computed from the Wilson score~\cite{wilson_probable_1927}. The gray lines correspond to the simulation results with added depolarising noise channel, as described in the text. The depolarising rate $p_{\mathrm{dpl}}$ is marked on every line.
    }
\end{figure}

\begin{figure}[t]
    \centering
    \includegraphics[width=0.48\textwidth]{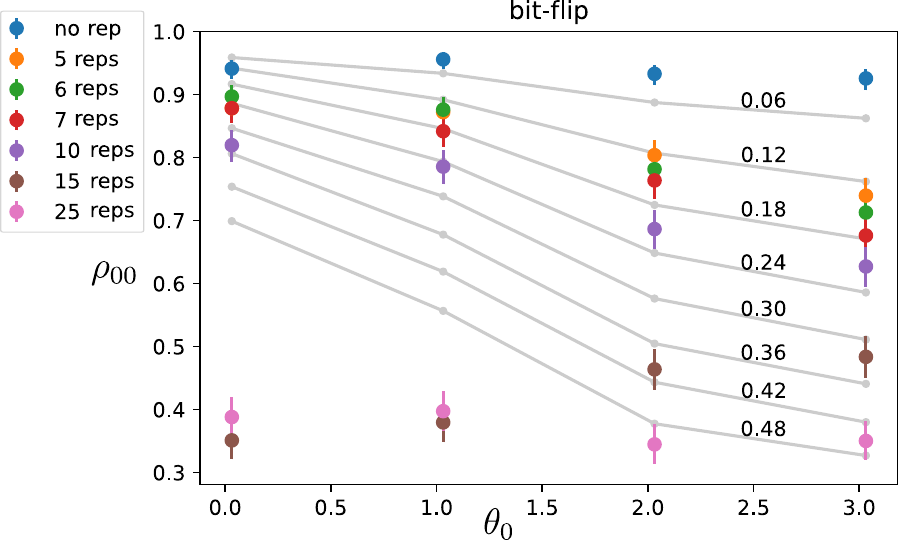}
    \caption{\label{fig:circuitSimBF}
        Comparing the data from the IonQ Aria-1 device with the results from the noisy simulations of quantum circuits, for the value of $\rho_{00}$, i.e., the diagonal element of the output logical state. The points with error bars correspond to the data, where certain numbers of redundant pairs of M{\o}lmer-S{\o}rensen gates have been added in the quantum circuit with native gates in order to tune the error rate. Each data point corresponds to an average over ${1000}$ shots. Error bars are ${95\%}$ confidence intervals computed from the Wilson score~\cite{wilson_probable_1927}. The gray lines correspond to the simulation results with added bit-flip noise channel, as described in the text. The bit-flip rate $p_{\mathrm{bf}}$ is marked on every line.
    }
\end{figure}

\section{the compiled circuits implemented on the IonQ Aria-1 quantum device}\label{sec:compiledCircuit}

The compiled circuits for the modified repetition code circuit shown in Fig.~2(a) in the main text that were run on the IonQ Aria-1 device are shown in Fig.~\ref{fig:circuitComplied}, corresponding to the data points in Fig.~3(a) in the main text except the black trangles pointing to the left. The four values of $\theta_0$ change the parameters of the GPi2 gates and M{\o}lmer-S{\o}rensen gates that are highlighted in red, where the first, second, third and fourth line inside the red blocks corresponds to the first, second, third and fourth value of $\theta_0$, respectively. The redundant pairs of M{\o}lmer-S{\o}rensen gates are marked in blue. Our definition of each repetition, as used in Fig.~3 in the main text, is that both pairs as circled in blue dotted lines are repeated. For example, for 5 repetitions, we will add 10 pairs of redundant M{\o}lmer-S{\o}rensen gates altogether. Note that the parameters of these redundant M{\o}lmer-S{\o}rensen gates are independent of the values of $\theta_0$.

\begin{figure}[t]
    \centering
    \includegraphics[width=0.48\textwidth]{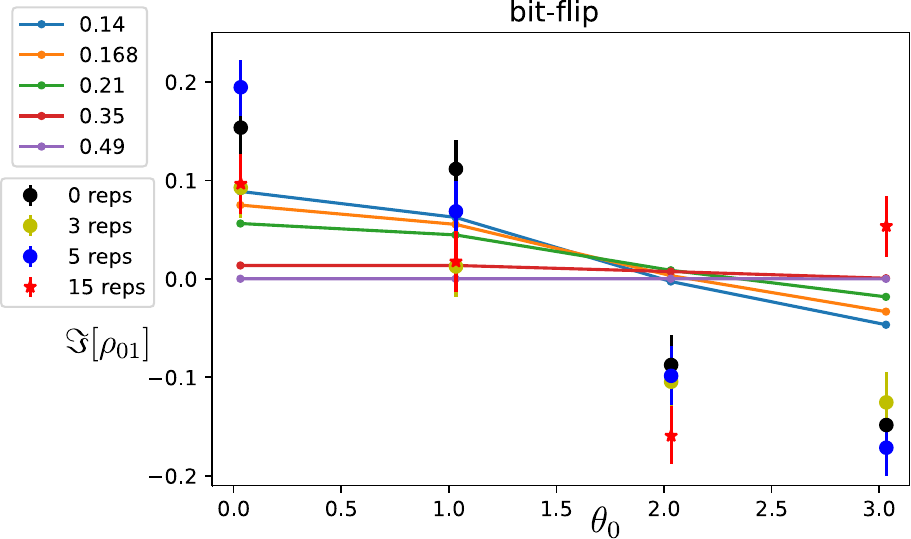}
    \caption{\label{fig:circuitFor3SimBF}
    Comparing the data from the IonQ Aria-1 device with the results from the noisy simulations of quantum circuits, for the value of $\Im[\rho_{01}]$, i.e., the imaginary part of the off-diagonal element of the output logical state. The quantum circuits are based on Fig.~2(b) in the main text and compiled to the native gates, having a very similar structure as Fig.~\ref{fig:circuitComplied}. In particular, the encoding part of the circuit (until the fourth group in Fig.~\ref{fig:circuitComplied}, also see the descriptions in the text), including the redundant M{\o}lmer-S{\o}rensen gate pairs, are the same as Fig.~\ref{fig:circuitComplied}. The points with error bars correspond to the data, where certain numbers of redundant pairs of M{\o}lmer-S{\o}rensen gates have been added in the quantum circuit with native gates in order to tune the error rate. Each data point corresponds to an average over ${1000}$ shots. Error bars are ${95\%}$ confidence intervals computed from the Wilson score~\cite{wilson_probable_1927}. The lines correspond to the simulation results with added bit-flip noise channel, as described in the text. The bit-flip rates $p_{\mathrm{bf}}$ are listed in the legend.
    }
\end{figure}

\section{Noisy simulations of quantum circuits}\label{sec:simCircuit}

Here in this section, we are going to show how to match the data from IonQ Aria-1 with noisy simulations of quantum circuits.

\begin{figure*}[h]
    \centering
    \includegraphics[width=0.75\textwidth]{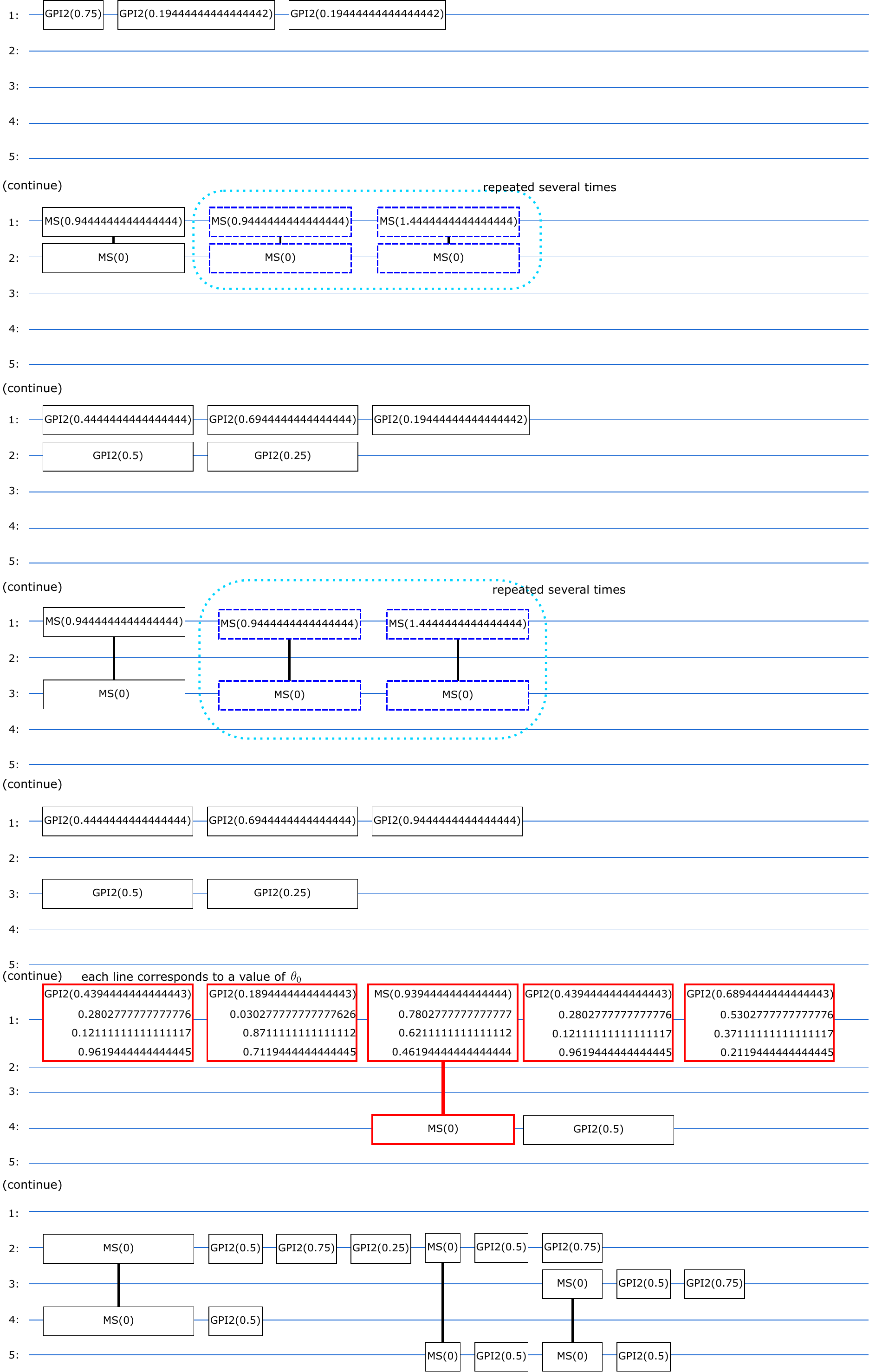}
    \caption{\label{fig:circuitComplied}
        The actual quantum circuits that were implemented on the IonQ Aria-1 quantum device. They correspond to the five-qubit syndrome measurement circuit (Fig.~2(a) in the main text).
        Redundant M{\o}lmer-S{\o}rensen gate pairs (corresponding to the identity channel in the ideal case) are inserted to tune the error rate. Note that the angles, after multiplied by $2\pi$, are in the unit of radius.}
\end{figure*}

We consider the quantum circuits made of native gates of the IonQ Aria-1 quantum device. For simulating $\rho_{00}$, the circuits are shown in Fig.~\ref{fig:circuitComplied}, except that we do not add in the redundant pairs of M{\o}lmer-S{\o}rensen gates as circled in blue dotted line, as on the simulator every quantum element is modelled as perfect. In order to mimic the effect of noise on the actual quantum device, we need to add in noise in the simulator. We replace the the redundant pairs of M{\o}lmer-S{\o}rensen gates with noise channels. We have chosen to use the predefined noise channels in Amazon Braket. Specifically, we have considered two models. The first one is the depolarising noise channel. After each of the two M{\o}lmer-S{\o}rensen gates that are followed by the blue circled blocks in Fig.~\ref{fig:circuitComplied}, we add a depolarising noise channel for each of the qubit involved with depolarising rate $p_{\mathrm{dpl}}$. That is, we have added altogether four depolarising noise channels. The second model is the bit-flip noise channel. The positions of these channels are the same as those in the first model. These channels are the bit-flip channel with bit-flip rate $p_{\mathrm{bf}}$.

Fig.~\ref{fig:circuitSimDpl} shows the results of noisy simulations with added depolarising noise, while Fig.~\ref{fig:circuitSimBF} is for the results of noisy simulations with added bit-flip noise. The slopes of the gray lines indicate that, for small $\theta_0$, depolarising noise fits better with the data, while for large $\theta_0$, bit-flip noise fits better. This is compatible with the fact that the quantum error correction code we work on before inserting the $R_y$ gate is designed to correct the bit-flip error. As seen in Fig.~\ref{fig:circuitSimBF}, along each gray line, the values of $\rho_{00}$ increases drastically as $\theta_0$ changes from the second value to the first value. The large population of $\rho_{00}$ at $\theta_0\approx0$ predicted by the noisy simulations is a result of bit-flip errors getting corrected by the repetition code. For the actual data obtained from the quantum device, however, there exists other sources of noise in addition to bit-flip noise. After the bit-flip errors are largely corrected by the code itself, the contributions from the remaining sources of error do not have a particular structure. Thus the change of $\rho_{00}$ from the first value to the second value of $\theta_0$ is better described by the depolarising noise channel, as seen in Fig.~\ref{fig:circuitSimDpl}.

We have also tried to match the data of $\Re[\rho_{01}]$ and $\Im[\rho_{01}]$ with the noisy simulations of quantum circuits. The quantum circuits before compilation for obtaining them by only measurements in the computational basis are shown in Fig.~2(b) in the main text. The compiled circuits are similar to the ones shown in Fig.~\ref{fig:circuitComplied} for measuring $\rho_{00}$, as described in details in the caption of Fig.~\ref{fig:circuitFor3SimBF}. We add bit-flip noise in the simulation, replacing the redundant pairs of M{\o}lmer-S{\o}rensen gates, same as the procedure described above. For $\Re[\rho_{01}]$, the noisy circuit simulation always has the result $\Re[\rho_{01}]=0$. For $\Im[\rho_{01}]$, the results are shown  in Fig.~\ref{fig:circuitFor3SimBF}. The data do not quantitatively match the noisy circuit simulation. This is partly because the relevant numbers are small, therefore it is difficult to distinguish from the noise level and reach a small error bar. Another possible reason is that the dependence of $\Im[\rho_{01}]$ on $\theta_0$ and $p_{\mathrm{bf}}$ is not strong enough to be resolved on the quantum device that has complicated noise sources. In particular, this dependence is not sensitive to the dominant noise source, which is from the repeated M{\o}lmer-S{\o}rensen gates whose errors are bit-flip like.

\section{Reshaping of noise with IQM Garnet}

We first transpile the modified three-qubit bit-flip repetition code circuit (Fig.~2(a) in the main text) into the native gate set of the IQM Garnet quantum device. Then we compare the hardware results from IQM Garnet with local simulations of the effects of increasing bit-flip, phase-flip, and depolarising errors on the transpiled circuit.

\subsection{Native Gate Set of IQM Garnet}

The native gate set of the IQM Garnet quantum device~\cite{abdurakhimov2024technology} includes the single-qubit PhaseRx (PRx) gates and two-qubit Controlled-Z gates. Here is the matrix representation of these gates: 
\begin{equation}
\text{PRx}(\theta, \phi) = \begin{pmatrix}
\cos(\theta/2) & -i e^{-i \phi} \sin(\theta/2) \\
-i e^{i \phi} \sin(\theta/2) & \cos(\theta/2)
\end{pmatrix}
\label{prx}
\end{equation}
and
\begin{equation}
CZ = 
\begin{pmatrix}
1 & 0 & 0 & 0 \\
0 & 1 & 0 & 0 \\
0 & 0 & 1 & 0 \\
0 & 0 & 0 & -1
\end{pmatrix},
\label{cz}
\end{equation}
where, in the computational basis, CZ gate flips the phase of the target qubit if the control qubit is in the $\ket{1}$ state.

\subsection{Mapping of CNOT to CZ Gates}
\begin{figure}[H]
    \centering
    \includegraphics[width=1\linewidth]{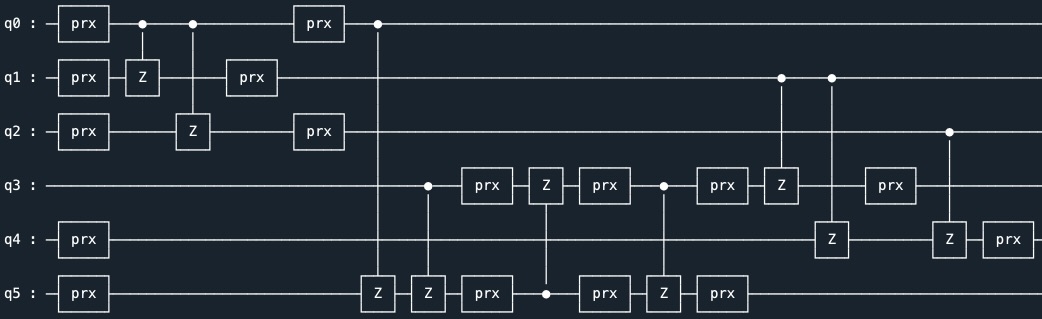}
    \caption{Quantum circuit for the three-qubit bit-flip repetition code with an added tunable single-qubit rotation gate $R_y(\theta_0)$ expressed with PRx and CZ gates.}
    \label{transpiled}
\end{figure}

We begin by running the three-qubit bit-flip repetition code with the added $R_y$ gate (main text Fig.~2(a) on the IQM Garnet quantum device. This allows us to obtain the transpiled version of the circuit directly from the returned results of the quantum device, see Fig.~\ref{transpiled}.
Note that this circuit representation remains the same for different values of $\theta_0$. Also, note that the hardware, which has 20 qubits in total, used qubits 3, 4, 8, 9, 13, and 14 to run the circuit, which we mapped and relabeled as qubits 0, 1, 2, 3, 4, and 5. Qubit 0 was mapped to qubit 9, qubit 1 to qubit 8, qubit 2 to qubit 4, qubit 3 to qubit 13, qubit 4 to qubit 3, and qubit 5 to qubit 14. Qubit 5 is an extra qubit added for transpilation purposes by the compiler, and it is very likely that SWAP operations were implemented due to qubit connectivity. Figure~\ref{map} is the qubit connectivity map for the IQM Garnet hardware, where we have identified in red the qubits used to implement our circuit:
\begin{figure}[H]
    \centering
    \includegraphics[width=0.75\linewidth]{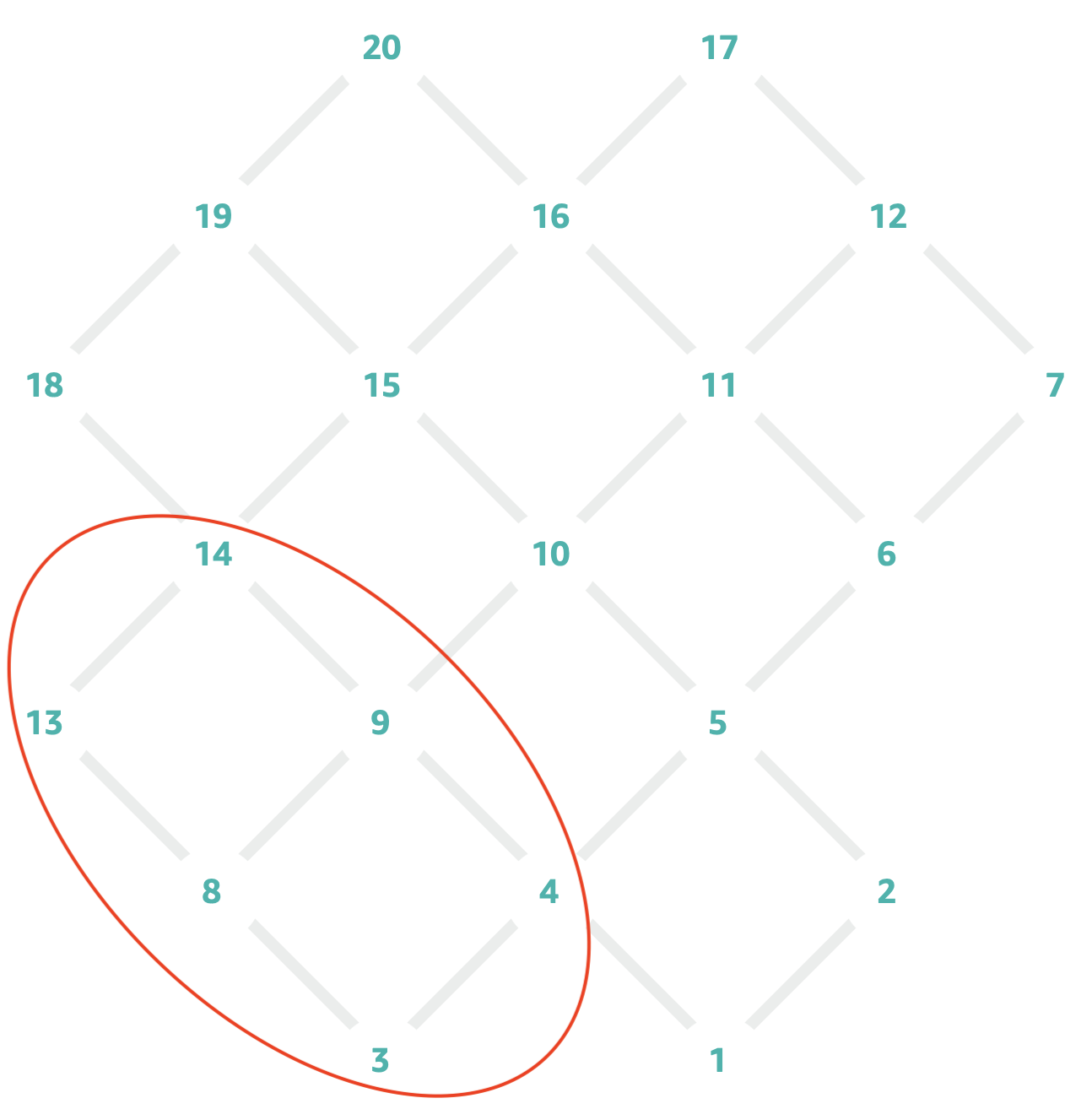}
    \caption{Qubit connectivity map for the IQM Garnet quantum device.}
    \label{map}
\end{figure}

\begin{figure}[H]
    \centering
    \includegraphics[width=0.48\textwidth]{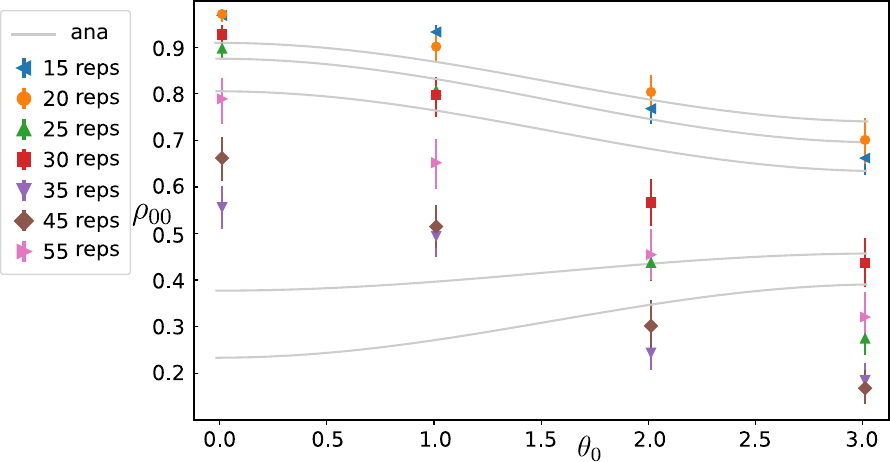}
    \caption{\label{fig:IQM}
    Measurement results obtained from the IQM Garnet quantum device averaged over 1000 shots, in comparison with the analytical circuit-agnostic model of Eq. (3) in the main text, where the grey lines correspond to different bit-flip error rates p. See the text for more details.
    }
\end{figure}

We can establish the following equivalence between the CNOT and CZ with PRx gates, up to a global phase:
\begin{align*}
&\text{CNOT} =  \text{H} \; \text{CZ} \; \text{H} \\
&= \frac{1}{\sqrt{2}} \begin{pmatrix}
1 & 1 \\
1 & -1
\end{pmatrix} 
\begin{pmatrix}
1 & 0 & 0 & 0 \\
0 & 1 & 0 & 0 \\
0 & 0 & 1 & 0 \\
0 & 0 & 0 & -1
\end{pmatrix}
\frac{1}{\sqrt{2}} \begin{pmatrix}
1 & 1 \\
1 & -1
\end{pmatrix} \\
&= \begin{pmatrix}
\cos(\pi/4) & -\sin(\pi/4) \\
\sin(\pi/4) & \cos(\pi/4)
\end{pmatrix} 
\begin{pmatrix}
\cos(\pi/2) & -i \sin(\pi/2) \\
-i \sin(\pi/2) & \cos(\pi/2)
\end{pmatrix} \\
& \hspace{0.5cm} \begin{pmatrix}
1 & 0 & 0 & 0 \\
0 & 1 & 0 & 0 \\
0 & 0 & 1 & 0 \\
0 & 0 & 0 & -1
\end{pmatrix} \\
& \hspace{0.5cm} \begin{pmatrix}
\cos(\pi/4) & -\sin(\pi/4) \\
\sin(\pi/4) & \cos(\pi/4)
\end{pmatrix} 
\begin{pmatrix}
\cos(\pi/2) & -i \sin(\pi/2) \\
-i \sin(\pi/2) & \cos(\pi/2)
\end{pmatrix} \\
&= R_y(\frac{\pi}{2})R_x(\pi) \; \text{CZ} \; R_y(\frac{\pi}{2})R_x(\pi) \\
&= \text{PRx}(\frac{\pi}{2}, \frac{\pi}{2})\text{PRx}(\pi, 0)  \; \text{CZ} \; \text{PRx}(\frac{\pi}{2}, \frac{\pi}{2})\text{PRx}(\pi, 0),
\end{align*}

where the Hadamard gate is $H = \frac{1}{\sqrt{2}} \begin{pmatrix} 1 & 1 \\ 1 & -1 \end{pmatrix}$.
It is difficult to precisely map the original circuit (main text Fig.~2(a)) to the compiled one in Fig.~\ref{transpiled}, but it is evident that the first two CZ gates, which are between qubits labeled as (0, 1) and (0, 2) and correspond to the qubits (9, 8) and (9, 4) on the physical hardware, encode the quantum information into the logical state, as they are associated with the first two CNOT gates in main text Fig.~2(a). These gates are applied before the gates related to the tunable parameter $\theta_0$. Later, we demonstrate the tunability of the noise by inserting different numbers of redundant pairs of CZ gates after each of the two encoding CZ gates. These pairs correspond to the identity operator in the noiseless case.

\subsection{Implementation of the Transpiled Quantum Circuit with extra CZ Gates on IQM Garnet} 

We have established that the two encoding CZ gates, corresponding to the first two CNOT gates of the repetition code with the inserted $R_y(\theta_0)$ gate, are located between qubits 9 and 8, and qubits 9 and 4 on the IQM Garnet hardware. The next step is to insert an increasing number of CZ gates after each of the two encoding CZ gates to amplify the noise that is naturally associated with the CZ gates. 

We initialize the circuit as follows: we choose the angle for $R_x$ from the circuit in main text Fig.~2(a) to be $\phi = \pi/9$. We will vary the angle of our tunable $R_y(\theta_0)$ gate, using the same values of $\theta_0$ as in the main text, specifically $\theta_0 = {0.01 \pi, 0.328 \pi, 0.647\pi, 0.965 \pi}$. We will run the compiled version of the circuit demonstrated in Fig.~\ref{transpiled} on the IQM Garnet device, and we are going to insert $15, 20, 25, 30, 35, 45,$ and $55$ redundant pairs of CZ gates after each of the two encoding CZ gates.

The primary source of noise in the implementation of the CZ gate is $1/f$ flux noise, originating from fluctuations in the magnetic flux through superconducting loops caused by material defects or environmental disturbances. This noise has been previously discussed as heavily affecting qubit phase. Additionally, parasitic ZZ interactions are also associated with the noise generated from imperfect CZ gate implementations~\cite{abdurakhimov2024technology}.

Fig.~\ref{fig:IQM} shows the measurement results obtained from running the three-qubit bit-flip repetition code with tunable single-gate operations, expressed using CZ and PRx gates, on the IQM Garnet quantum device. Each data point represents an average over 1000 shots, with error bars indicating 95$\%$ confidence intervals computed using the Wilson score. We varied the values of angle $\theta_0$ and the number of inserted redundant CZ gate pairs. Adding a small number of redundant CZ gate pairs reveals the dependence of $\rho_{00}$ on $\theta_0$. This includes 15, 20, 25, and 35 repetitions, corresponding to blue triangles, orange circles, green triangles, red squares, and purple triangles, respectively, on the plot. As $\theta_0$ increases from $0$ to $\pi$, $\rho_{00}$ consistently decreases. This indicates that we have successfully managed to tune the channel in a predictable way and that this tunability depends on both the profile and level of noise. The results obtained from the hardware demonstrate that the native noise of the device enables a useful feature. We observe that while the decrease in $\rho_{00}$ with increasing $\theta_0$ holds true for a larger number of redundant CZ gate pairs, the overall value of $\rho_{00}$ starts to shift upward past 35 repetitions: the overall value of $\rho_{00}$ for each $\theta_0$ is higher for 45 repetitions (represented by brown diamonds) than for 35, and it is also higher for 55 repetitions (represented by pink triangle) than for 45. We superimposed the plot of the analytical, circuit-agnostic model, main text Eq.~(3) in Fig.\ref{fig:IQM}. We observe that the analytical model does not match the results from the hardware. As the CZ gate is unlikely to have bit-flip noise, we do expect that there is no match.

\subsection{Simulations of the Effects of the Increasing Bit-Flip, Phase-Flip and depolarising Errors on the Transpiled Circuit}

\begin{figure}[H]
    \centering
    \includegraphics[width=1\linewidth]{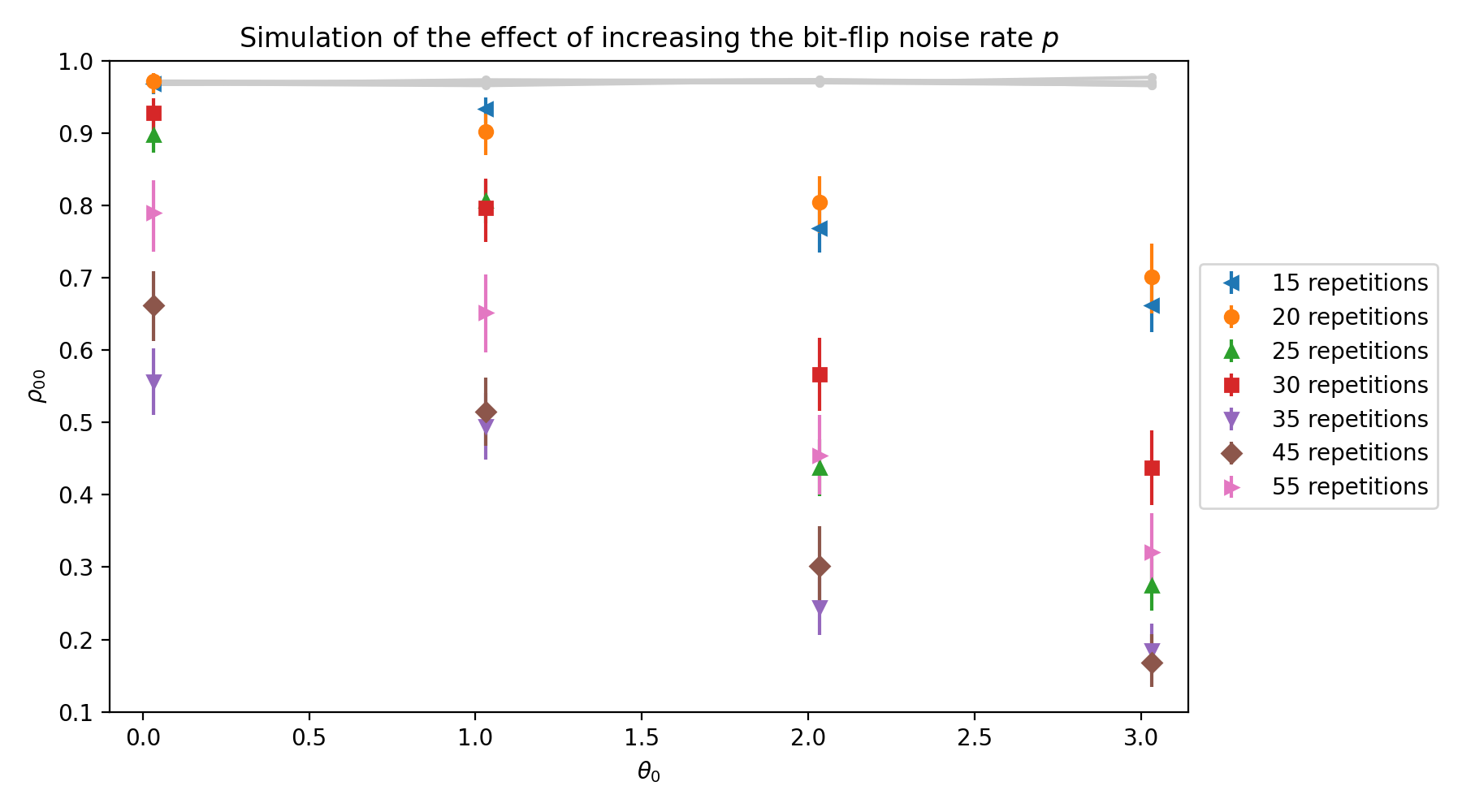}
    \caption{Comparison of the data obtained from the IQM Garnet quantum device with the simulations results of quantum circuits including bit-flip noise.}
    \label{bit flip}
\end{figure}

\begin{figure}[H]
    \centering
    \includegraphics[width=1\linewidth]{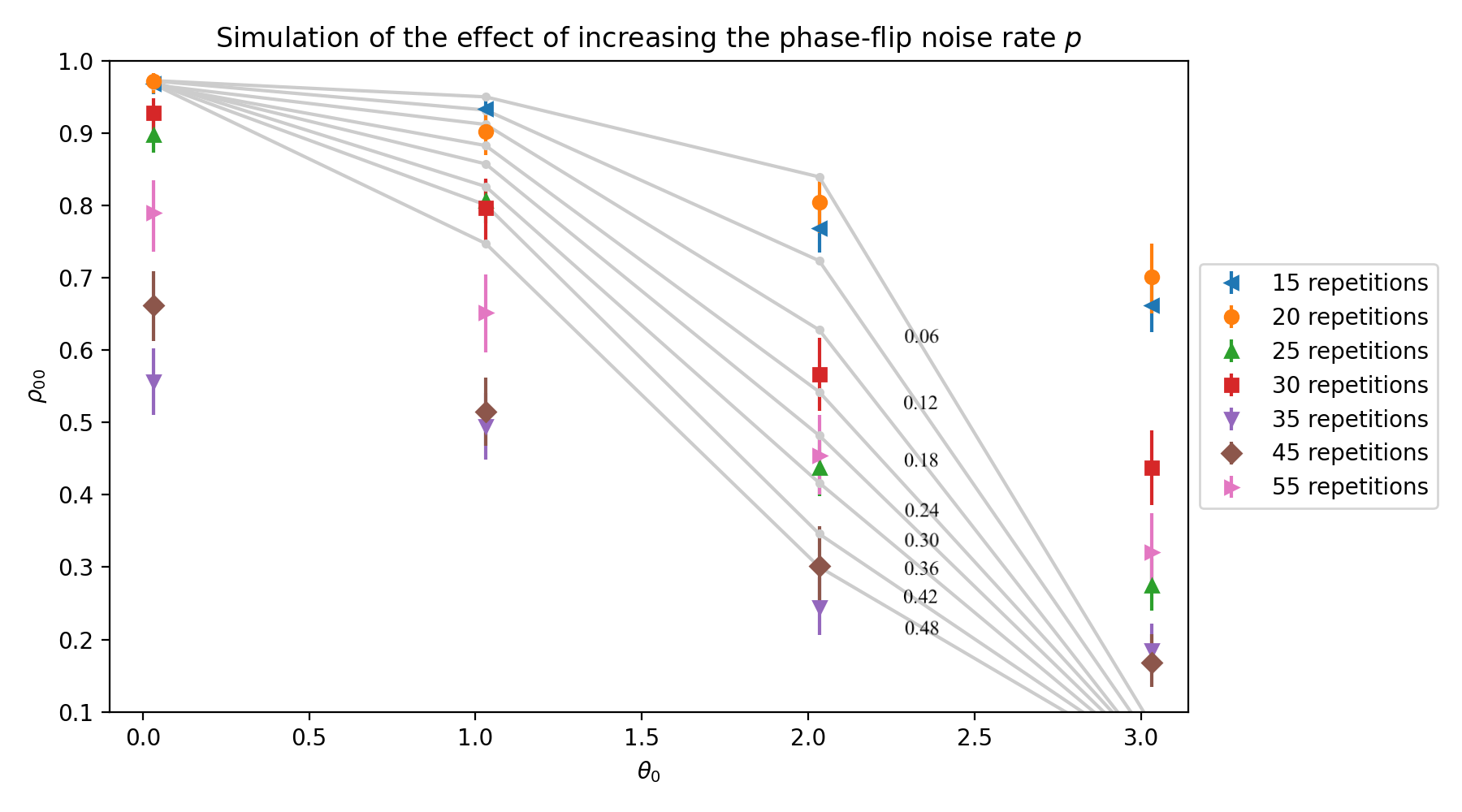}
    \caption{Comparison of the data obtained from the IQM Garnet quantum device with the simulations results of quantum circuits including phase-flip noise.}
    \label{phase flip}
\end{figure}

\begin{figure}[H]
    \centering
    \includegraphics[width=1\linewidth]{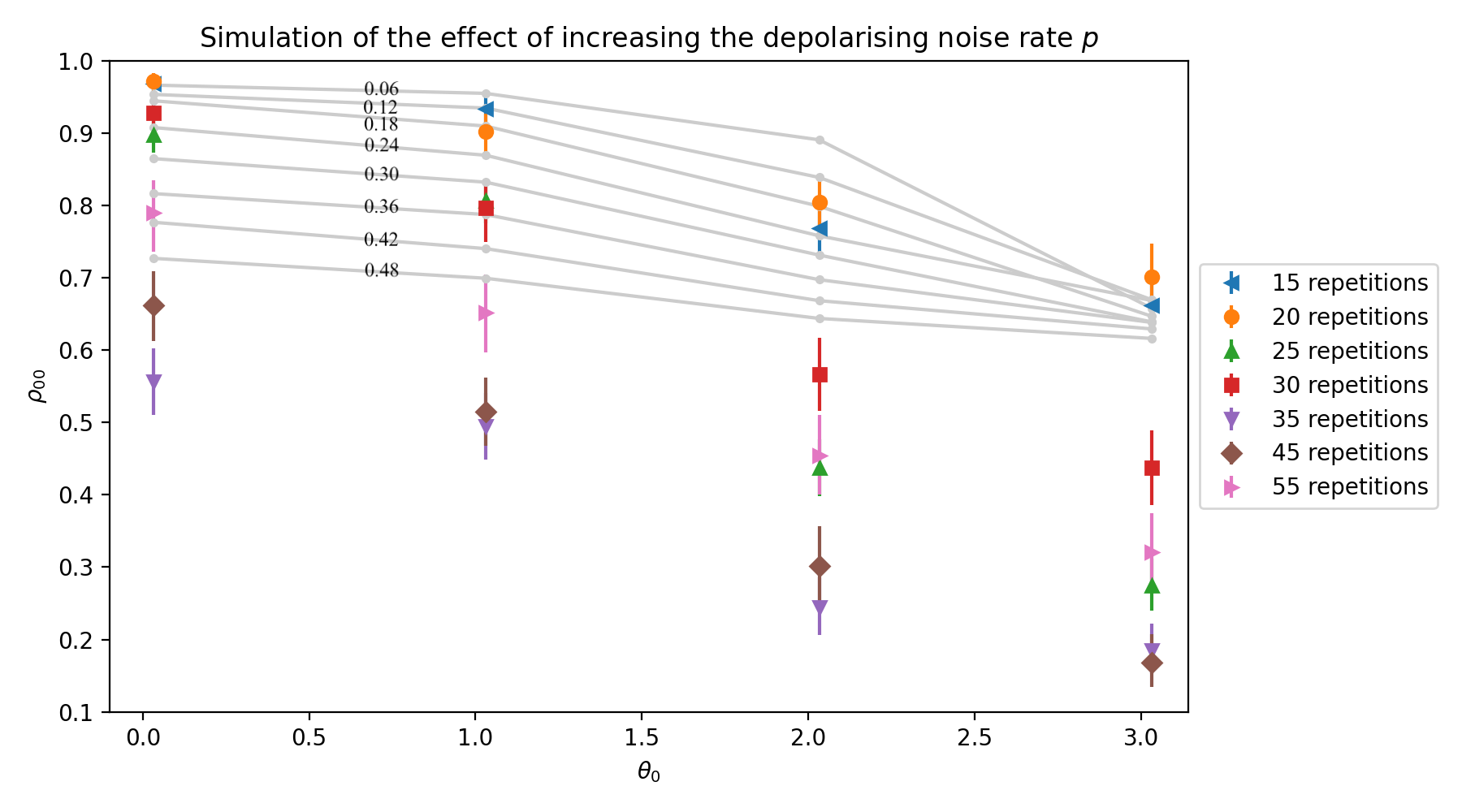}
    \caption{Comparison of the data obtained from the IQM Garnet quantum device with the simulations results of quantum circuits including depolarising noise.}
    \label{depolarising}
\end{figure}

To further investigate the noise profile of CZ gates, we simulated the effects of increasing noise on the three-qubit bit-flip repetition code with a tunable $R_y(\theta_0)$ gate constructed using CZ and PRx gates. In the simulator, adding redundant pairs of CZ gates does not alter the output logical state, as every quantum operation is modeled as perfect. Therefore, instead of inserting specific repetitions of the CZ gate, we added predefined noise channels in Amazon Braket to mimic the effect of noise on the actual quantum device. We simulated the effect of increasing bit-flip noise, phase-flip noise, and depolarising noise by directly inserting these noise models into the circuit. To do this, we added the noise channels one at a time after each of the two encoding CZ gates: a CZ gate between qubits (9, 8) and a CZ gate between qubits (9, 4). We inserted two noise channels after each encoding CZ gate, resulting in four noise channels in total. The error rates were set to $p={0.06, 0.12, 0.18, 0.24, 0.30, 0.36, 0.42, 0.48}$, corresponding to the gray lines in Figures \ref{bit flip}, \ref{phase flip}, and \ref{depolarising}. The simulations show the measurement results averaged over 10,000 shots. We compared these simulations with the hardware measurements to determine which noise model is most closely associated with the imperfect implementation of the CZ gate.

The results of the noisy simulations with added bit-flip noise are shown in Fig. \ref{bit flip}, while Fig. \ref{phase flip} displays the results for phase-flip noise, and Fig. \ref{depolarising} illustrates depolarising noise. As seen in Fig. \ref{bit flip}, the bit-flip noise model does not fit the hardware results. We can deduce that the noise model associated with the CZ gate does not match bit-flip error noise model. 

Turning to the plot of noisy simulations with phase-flip errors (Fig. \ref{phase flip}), we observe a better fit with the hardware data compared to the bit-flip error case. However, the slopes of the gray lines, which correspond to different phase-flip error rates, are steeper than the slopes of the data points from the IQM Garnet quantum device measurements. This indicates that the simulated phase-flip noise model does not fully match the hardware data, suggesting that the imperfect implementation of CZ gates generates a noise model which is not only phase-flip noise model but also a mix of other types of noise. Nonetheless, we confirm that phase-flip noise model, unlike bit-flip noise model, is a significantly better match with the noise of the CZ gate. 

We interpret the simulations of increasing depolarising noise shown in Fig. \ref{depolarising} similarly. The plot of the depolarising noise model shows similarities with the hardware measurement results, suggesting that the noise generated by the imperfect realization of the CZ gate also includes depolarising noise model among other types. However, the slopes of the gray lines are shallower than the slopes of the hardware measurement data, indicating that depolarising noise alone does not fully account for the observed data and is not the only type of noise model present in the channel.